\documentclass[twocolumn,twocolappendix]{aastex631}

\usepackage{subfigure}
\usepackage{graphbox}
\usepackage{comment}
\usepackage{threeparttable}
\usepackage{color}
\usepackage{CJK}
\usepackage{placeins}

\graphicspath{{fig/}}
\DeclareGraphicsExtensions{.pdf,.png,.jpg}

\definecolor{pltred}{RGB}{232,55,32}
\definecolor{pltgreen}{RGB}{59,166,33}
\definecolor{pltblue}{RGB}{32,11,245}

\begin{document}
\begin{CJK*}{UTF8}{gbsn}

\title{Galaxy Metallicity Gradients in the Epoch of Reionization from the FIRE-2 Simulations}

\correspondingauthor{Xin Wang}
\email{xwang@ucas.ac.cn}


\author[0009-0005-8170-5153]{Xunda Sun}
\affil{School of Astronomy and Space Science, University of Chinese Academy of Sciences (UCAS), Beijing 100049, China}

\author[0000-0002-9373-3865]{Xin Wang}
\affil{School of Astronomy and Space Science, University of Chinese Academy of Sciences (UCAS), Beijing 100049, China}
\affil{National Astronomical Observatories, Chinese Academy of Sciences, Beijing 100101, China}
\affil{Institute for Frontiers in Astronomy and Astrophysics, Beijing Normal University,  Beijing 102206, China}

\author[0000-0001-6115-0633]{Fangzhou Jiang}
\affil{Kavli Institute for Astronomy and Astrophysics, Peking University, Beijing 100871, China}

\author[0000-0001-5356-2419]{Houjun Mo}
\affil{Department of Astronomy, University of Massachusetts, Amherst, MA 01003, USA}

\author[0000-0001-6947-5846]{Luis C. Ho}
\affil{Kavli Institute for Astronomy and Astrophysics, Peking University, Beijing 100871, China}
\affil{Department of Astronomy, School of Physics, Peking University, Beijing 100871, China}

\author{Qianqiao Zhou}
\affil{School of Astronomy and Space Science, University of Chinese Academy of Sciences (UCAS), Beijing 100049, China}

\author{Xiangcheng Ma}
\affil{Department of Astronomy and Theoretical Astrophysics Center, University of California Berkeley, Berkeley, CA 94720, USA}

\author[0000-0003-1718-6481]{Hu Zhan}
\affil{National Astronomical Observatories, Chinese Academy of Sciences, Beijing 100101, China}
\affil{Kavli Institute for Astronomy and Astrophysics, Peking University, Beijing 100871, China}


\author[0000-0003-0603-8942]{Andrew Wetzel}
\affil{Department of Physics and Astronomy, University of California, Davis, CA, USA 95616}

\author{Russell L. Graf}
\affil{Department of Physics and Astronomy, University of California, Davis, CA, USA 95616}

\author[0000-0003-3729-1684]{Philip F. Hopkins}
\affil{TAPIR, Mailcode 350-17, California Institute of Technology, Pasadena, CA 91125, USA}

\author[0000-0002-1666-7067]{Du\v{s}an Kere\v{s}}
\affil{Department of Physics, Center for Astrophysics and Space Sciences, University of California San Diego, 9500 Gilman Drive, La Jolla, CA 92093, USA}

\author{Jonathan Stern}
\affil{School of Physics \& Astronomy, Tel Aviv University, Tel Aviv 69978, Israel}

\begin{abstract}
We employ the high-redshift suite of FIRE-2 cosmological hydrodynamic zoom-in simulations to investigate the evolution of gas-phase metallicity radial gradients in galaxies in the epoch of reionization (EoR). Our sample consists of 22 galaxies spanning the redshift range $z \sim 10-5$. 
We find that galaxies at $z\sim10$ exhibit a median metallicity gradient of $-0.15\,\mathrm{dex\cdot kpc^{-1}}$ with substantial scatter, which gradually flattens to $-0.1\,\mathrm{dex\cdot kpc^{-1}}$ at $z\sim6$, accompanied by a reduction in scatter. 
In the EoR, metallicity gradients correlate positively with stellar mass: more massive galaxies display flatter gradients with smaller scatter, broadly consistent with recent JWST observations. 
At fixed stellar mass, galaxies with higher star formation rate (SFR) exhibit steeper negative gradients, while specific SFR (sSFR) shows a strong anti-correlation with gradient slope. 
Because EoR galaxies in FIRE-2 generally lack significant rotational support, we adopt the ratio of peak-to-peak velocity shear to twice the velocity dispersion ($\Delta v/2\sigma$) as a proxy for the strength of gas flows. We find a strong positive correlation between metallicity gradients and $\Delta v/2\sigma$: galaxies with lower $\Delta v/2\sigma$ (i.e., weaker gas flows) tend to exhibit steeper negative gradients. Furthermore, galaxies with steeper gradients display higher central SFR surface densities, suggesting localized star formation with inefficient interstellar medium mixing that drives inside-out chemical enrichment in galaxy evolution in the early Universe. 

\end{abstract}

\keywords{Hydrodynamical simulations --  Galaxy evolution -- Galaxy formation -- Interstellar medium -- Metallicity -- High-redshift galaxies -- Reionization}

\section{Introduction} \label{sec:intro}


The spatial distribution of metals provides a key tracer of galaxy formation and evolution.
Metals, primarily produced by stellar nucleosynthesis \citep[e.g.][]{Nomoto2013}, are injected into the interstellar medium (ISM) through stellar winds and supernova explosions \citep[e.g.][]{Wangk2023}.
Galaxy metallicity is commonly discussed in two components: gas-phase metallicity, which tracks the present chemical state of the ISM and is closely tied to ongoing star formation and feedback \citep[e.g.][]{Tremonti2004, Lee2006}, and stellar metallicity, which records the time-integrated enrichment history over cosmic time \citep[e.g.][]{Gallazzi2005, Kirby2013}.

Previous studies have revealed a strong correlation between stellar mass and gas-phase metallicity in galaxies, commonly referred to as the mass-metallicity relation (MZR), which indicates that more massive galaxies tend to be more metal-rich in both the local Universe \citep[e.g.][]{Tremonti2004, Zahid2012, Andrews2013} and at cosmic noon \citep[e.g.][]{Erb2006, Maiolino2008, Sanders2015, Ma2016, Sanders2021, He2024}.
Recent efforts have extended the study to the epoch of reionization (EoR) \citep{Nakajima2023, Curti2024, Marszewski2024, LiS2025}.
A secondary dependence on star formation rate (SFR) has also been proposed, giving rise to the fundamental metallicity relation (FMR) \citep{Mannucci2010, Bothwell2013, Marszewski2025}.
While the MZR and FMR describe global metallicity scaling relations, understanding the internal chemical structure of galaxies requires spatially resolved measurements.
A common metric is the metallicity gradient, quantified by fitting a linear slope to the radial metallicity profile \citep[e.g.][]{Jones2015, Wang2017, Wang2020, Wang2022}.

The evolution of metallicity gradients has been widely explored observationally at low redshift. 
Since \citet{Searle1971}, it has been well established that the central regions of nearby galaxies are more metal-rich than their outskirts, yielding negative radial metallicity gradients \citep[e.g.][]{Zaritsky1994, vanZee1998}. 
Observational studies also suggest that gradients vary systematically with galaxy mass and morphology \citep[e.g.][]{Menguiano2016, Belfiore2017, Carton2018, Poetrodjojo2021, LiTie2025, Ju2025, Lyu2025, Khoram2025}. 
Mergers and interactions can further redistribute gas-phase metals and, in some cases, flatten or disturb gradients by driving gas inflows and enhanced star formation \citep[e.g.][]{Rupke20101, Rupke2010ApJ1255R, Cortijo2017, Porter2022, SunW2024}. 
The metal distribution reflects the combined effects of gas accretion, stellar feedback, mergers, active galactic nuclei (AGN) feedback, and metal transport/mixing processes \citep[e.g.][]{Mo2010, King2015, Thielemann2017, Wangk2023, Nakajima2024, He2024S, Peng2025P, Mo20241, Chen20242, Chen20253}. 
Numerical simulations further report correlations between gradients and galaxy properties: several works find that galaxies with higher specific SFR (sSFR) and lower mass tend to exhibit flatter gradients at $z\sim0-3$ \citep[e.g.][]{Stott2014, Wuyts2016, Sillero2017, Ma2017_FIRE_gra, Hemler2021, Tissera2022}, while steep negative gradients are more common in systems with significant rotational support \citep[e.g.][]{Bellardini2022, Graf2025, Sun2024_z04_3, Liang2025}. 
However, these trends can become less predictive in the presence of strong feedback, which can flatten gradients and drive rapid time variability \citep{Sharda2021, Hemler2021}.

Several studies have reported the presence of positive metallicity gradients in some galaxies during cosmic noon and local universe \citep{Cresci2010, Wang2019, Maiolino2019, Kewley2019}, suggesting that metal enrichment may not always proceed from the center outward.
\citet{Gibson2013} used cosmological hydrodynamical simulations to compare two feedback implementations, finding that the enhanced-feedback model ``MaGICC'' produces significantly flatter metallicity gradients than the more conservative one ``MUGS'' \cite[see also:][]{Luo2024}.
Recent studies suggest that a single linear metallicity gradient may not capture the full complexity of metal distributions. 
Spatially resolved observations and simulations report azimuthal variations and departures from simple radial trends \citep[e.g.][]{Ho2015, Ho2017, Kreckel2019, Bellardini2021, Bellardini2022, Baker2025}, and several works propose higher-order, multiscale statistics to quantify metal inhomogeneities \citep[e.g.][]{Krumholz2018, Metha2021}.
Nevertheless, in most cases a linear fit remains the simplest approach for gradient analysis, while also facilitating a direct comparison with high-z observations given the level of spatial detail currently available.

Recently, new data from JWST have opened up the possibility of observationally probing metallicity gradients in EoR galaxies for the first time.
Recent studies \citep{Arribas2024_z7, Vallini2024_z7, Venturi2024_z6-8, Tripodi2024_z4-10} have provided initial constraints on metallicity gradients at $z>6$, revealing a diverse range of gradient behaviors in EoR galaxies.
Most results suggest that EoR galaxies tend to exhibit modestly negative gradients around $-0.05\, \mathrm{dex\cdot kpc^{-1}}$, and a subset of observations have also reported the presence of positive gradients.
\citet{Garcia_EAGLE_TNG, Garcia_SPICE} present results from TNG, Illustris, EAGLE, SPICE and Thesan simulations in the EoR, showing that most simulated galaxies at this epoch exhibit strong negative gradients accompanied by substantial scatter.
In the early stages of galaxy evolution, galaxies exhibit evolutionary behaviors that differ markedly from those at low redshift, with feedback processes playing a correspondingly different role \citep[e.g.][]{Jin2024a, Bassini20231, Mo20241}.
\citet{SunGC2026} further argued that the gas in many EoR galaxies is turbulence-dominated rather than supported by stable disk dynamics, leading star formation to proceed in localized, clumpy dense regions.
However, studies focusing on metallicity gradients in the EoR remain limited.
Only a few simulation suites have explored this regime with sufficient resolution and sample size, leaving the physical drivers of early metallicity gradients still incompletely understood.

The Feedback In Realistic Environments simulations (FIRE) \footnote{See also: \url{https://fire.northwestern.edu/}} have been widely utilized to investigate the evolution of metallicity gradients under a broad range of physical conditions. These include three-dimensional abundance structures \citep{Bellardini2021, Bellardini2022, Graf2025}, the impact of stellar feedback at cosmic noon \citep{Ma2017_FIRE_gra, Sun2024_z04_3}, gas inflows and turbulent mixing \citep{Graf2024}, and azimuthal variations associated with spiral structures \citep{Orr2023}, among others.
Building on these studies, we conduct a detailed analysis of gas-phase metallicity gradients in galaxies in the EoR using the high-redshift suite of the FIRE-2 project. This enables us to examine how chemical structures form and evolve in the earliest stages of galaxy assembly.

This paper is organized as follows. In Section~\ref{sec:method}, we describe the simulation setup and the methods used to define galaxies and compute their physical properties. Section~\ref{sec:result} presents our main results, including comparisons with observational data and other cosmological simulations. We conclude in Section~\ref{sec:conclusion} with a summary of our key findings and their implications for early galaxy evolution.

\section{Methodology} \label{sec:method}
\subsection{Simulation}

\begin{table*}[!htb]
\caption{Simulation details.}
\centering
\begin{threeparttable}
\label{tab:galapro}
    \centering
    \setlength{\tabcolsep}{12pt} 
\begin{tabular*}{0.9\linewidth}{@{}ccccccccc@{}}
\hline
Name & $M_{\rm halo}$ & $m_{\rm baryon}$ & $m_{\rm dm}$ & $\epsilon_{\rm star}$ & $\epsilon_{\rm dm}$ & $\epsilon_{\rm gas, min}$ & Reference \\
  & (M$_\odot$) & (M$_\odot$) & (M$_\odot$) & ({\rm pc}) & ({\rm pc}) &({\rm pc}) & \\
\hline
 \texttt{z5m09a} & $2.4\times10^{9}$ & 119 & 650 & 0.7 & 10 & 0.14 & \cite{Ma2018_FIRE} \\
 \texttt{z5m09b} & $3.9\times10^{9}$ & 119 & 650 & 0.7 & 10 & 0.14 & \cite{Ma2018_FIRE} \\
 \texttt{z5m10a} & $6.6\times10^{9}$ & 119 & 650 & 0.7 & 10 & 0.14 & \cite{Ma2020_FIRE} \\
 \texttt{z5m10b} & $1.2\times10^{10}$ & 954 & 5200 & 1.4 & 21 & 0.28 & \cite{Ma2018_FIRE} \\
 \texttt{z5m10c} & $1.3\times10^{10}$ & 954 & 5200 & 1.4 & 21 & 0.28 & \cite{Ma2018_FIRE} \\
 \texttt{z5m10d} & $1.9\times10^{10}$ & 954 & 5200 & 1.4 & 21 & 0.28 & \cite{Ma2018_FIRE} \\
 \texttt{z5m10e} & $2.6\times10^{10}$ & 954 & 5200 & 1.4 & 21 & 0.28 & \cite{Ma2018_FIRE} \\
 \texttt{z5m10f} & $3.3\times10^{10}$ & 954 & 5200 & 1.4 & 21 & 0.28 & \cite{Ma2018_FIRE} \\
 \texttt{z5m11a} & $4.2\times10^{10}$ & 954 & 5200 & 1.4 & 21 & 0.28 & \cite{Ma2018_FIRE} \\
 \texttt{z5m11b} & $4.0\times10^{10}$ & 891 & 4900 & 1.4 & 21 & 0.28 & \cite{Ma2018_FIRE} \\
 \texttt{z5m11i} & $5.2\times10^{10}$ & 891 & 4900 & 1.4 & 21 & 0.28 & \cite{Ma2020_FIRE} \\
 \texttt{z5m11c} & $7.6\times10^{10}$ & 891 & 4900 & 1.4 & 21 & 0.28 & \cite{Ma2020_FIRE} \\
 \texttt{z5m11h} & $1.0\times10^{11}$ & 7100 & $3.9\times10^4$ & 2.1 & 42 & 0.42 & \cite{Ma2019_FIRE} \\
 \texttt{z5m11d} & $1.4\times10^{11}$ & 7100 & $3.9\times10^4$ & 2.1 & 42 & 0.42 & \cite{Ma2018_FIRE} \\
 \texttt{z5m11g} & $2.0\times10^{11}$ & 7100 & $3.9\times10^4$ & 2.1 & 42 & 0.42 & \cite{Ma2019_FIRE} \\
 \texttt{z5m11e} & $2.5\times10^{11}$ & 7100 & $3.9\times10^4$ & 2.1 & 42 & 0.42 & \cite{Ma2018_FIRE} \\
 \texttt{z5m11f} & $3.1\times10^{11}$ & 7100 & $3.9\times10^4$ & 2.1 & 42 & 0.42 & \cite{Ma2019_FIRE} \\
 \texttt{z5m12a} & $4.5\times10^{11}$ & 7100 & $3.9\times10^4$ & 2.1 & 42 & 0.42 & \cite{Ma2018_FIRE} \\
 \texttt{z5m12e} & $5.0\times10^{11}$ & 7100 & $3.9\times10^4$ & 2.1 & 42 & 0.42 & \cite{Ma2019_FIRE} \\
 \texttt{z5m12d} & $5.7\times10^{11}$ & 7100 & $3.9\times10^4$ & 2.1 & 42 & 0.42 & \cite{Ma2019_FIRE} \\
 \texttt{z5m12c} & $7.9\times10^{11}$ & 7100 & $3.9\times10^4$ & 2.1 & 42 & 0.42 & \cite{Ma2019_FIRE} \\
 \texttt{z5m12b} & $8.7\times10^{11}$ & 7100 & $3.9\times10^4$ & 2.1 & 42 & 0.42 & \cite{Ma2018_FIRE} \\
\hline
\end{tabular*}
\begin{tablenotes}
\item \hspace{-13pt}Parameters describing the initial conditions for our simulations (units are physical):
\item Name: simulation designation.
\item $M_{\rm halo}$: approximate mass of the main halo at $z = 5$.
\item $m_{\rm baryon}$, $m_{\rm dm}$: initial masses of baryonic (gas or star) and dark-matter particles.
\item $\epsilon_{\rm star}$, $\epsilon_{\rm dm}$: force softening (Plummer equivalent) for star and dark-matter particles.
\item $\epsilon_{\rm gas, min}$: minimum adaptive force softening for gas cells.
\item Reference: where the simulation is first presented.
\item All simulations use cosmological parameters: 
        \item \hspace{13pt}P (`Planck': $\Omega_{\rm m}$ = 0.31, $\Omega_\Lambda$ = 0.69, $\Omega_{\rm b}$ = 0.0458, $h$ = 0.68, $\sigma_8$ = 0.82, $n_{\rm s}$ = 0.97).
\end{tablenotes}
\end{threeparttable}
\end{table*}

The simulations used in this study derive from the output of the high-redshift suite of the FIRE-2 cosmological hydrodynamic simulations in \citet{Ma2018_FIRE, Ma2019_FIRE, Ma2020_FIRE}. All simulations were performed using the \textsc{gizmo} code in its meshless finite-mass (MFM) mode \citep{Hopkins2015_FIRE}  and the FIRE-2 physics model \citep{Hopkins2018_FIRE2INTRO}.
Our sample includes 22 zoom-in simulations of galaxies: \texttt{z5m09a-b}, \texttt{z5m10a-f}, \texttt{z5m11a-i}, and \texttt{z5m12a-e}. These galaxies are evolved down to $z = 5$, spanning final main halo masses ranging from $M_{\mathrm{h}} \sim 10^9$ to $10^{12}\,\mathrm{M}_\odot$.
We analyze 11 snapshots for each galaxy uniformly spaced in cosmic time over the redshift interval $z \sim 10$-5, resulting in 242 snapshots in total.
For these galaxies, the initial mass of baryonic and dark matter particles are $m_{\mathrm{baryon}}=119-7100\,\mathrm{M}_\odot$ and $m_{\mathrm{dm}}=650-3.9\times10^4\,\mathrm{M}_\odot$.
Gravitational force softening lengths (Plummer-equivalent) span $\epsilon_{\mathrm{star}} = 0.7-2.1\,\mathrm{pc}$ for star particles and $\epsilon_{\mathrm{dm}} = 10-42\,\mathrm{pc}$ for dark matter. 
Gas follows an adaptive softening scheme, with a minimum resolution reaching $\epsilon_{\mathrm{gas, min}} = 0.14-0.42\,\mathrm{pc}$.
All runs adopt a standard flat $\Lambda \mathrm{CDM}$ cosmology with \emph{Planck2015} \citep{Planck2016} cosmological parameters; see Table~\ref{tab:galapro} for a summary of galaxy properties.

The FIRE-2 simulations track the abundances of 11 different elements (H, He, C, N, O, Ne, Mg, Si, S, Ca, Fe,  following \citet{Wiersma2009_elem}), and span a broad temperature range from 10 to $10^{10}\,\mathrm{K}$, enabling accurate modeling of gas thermodynamics across different phases of the ISM. 
Multiple stellar feedback processes are implemented, including radiative feedback, Type Ia and Type II supernovae, and stellar winds from O/B and AGB stars. These mechanisms drive metal mixing and energy transfer within the ISM.
These channels are coupled explicitly and locally to the surrounding gas, leading to a comparatively strong and time-variable impact of feedback.
The stellar evolution model in the simulations adopts a Kroupa initial mass function (IMF; \citet{Kroupa2001}), with feedback rates and energetics derived from \textsc{starburst99} \citep{Leitherer1999}.
The simulations also include a subgrid turbulent diffusion model to account for unresolved metal mixing through local gas turbulence, allowing chemical elements to diffuse between neighboring resolution elements.
For more discussion of its impact on the differences relative to other simulations, see Section~\ref{subsec:gradtoz}.
These FIRE-2 simulations do not model feedback from AGN.

Further details on the high-redshift simulation suite are presented in \citet{Ma2018_FIRE, Ma2019_FIRE, Ma2020_FIRE}, while the overall FIRE-2 framework is described in \citet{Hopkins2018_FIRE2INTRO}. A summary of the publicly released FIRE-2 data is provided in \citet{Wetzel2023}.

\begin{figure*}
\begin{minipage}{0.6\linewidth}
 \centering
\subfigure{
 \centering
 \includegraphics[align=c,width=0.9\linewidth]{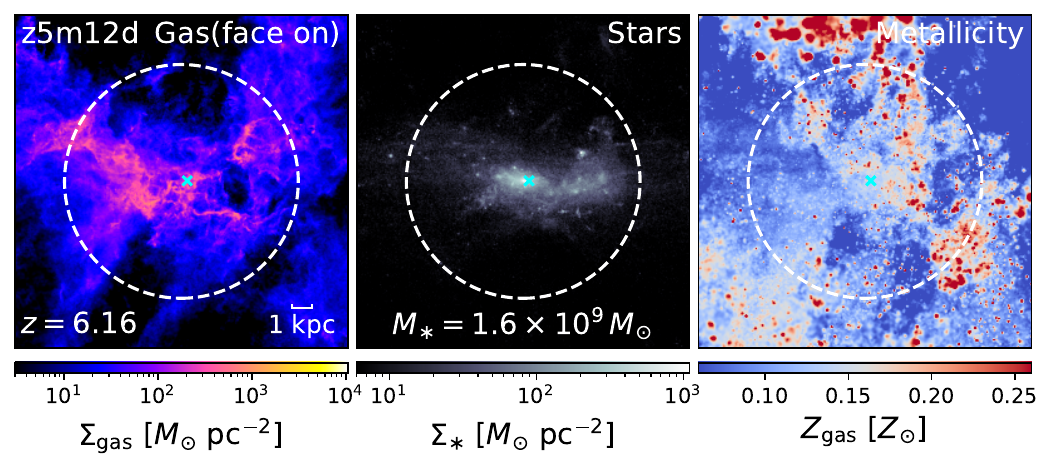}}
 \subfigure{
 \centering
 \includegraphics[align=c,width=0.9\linewidth]{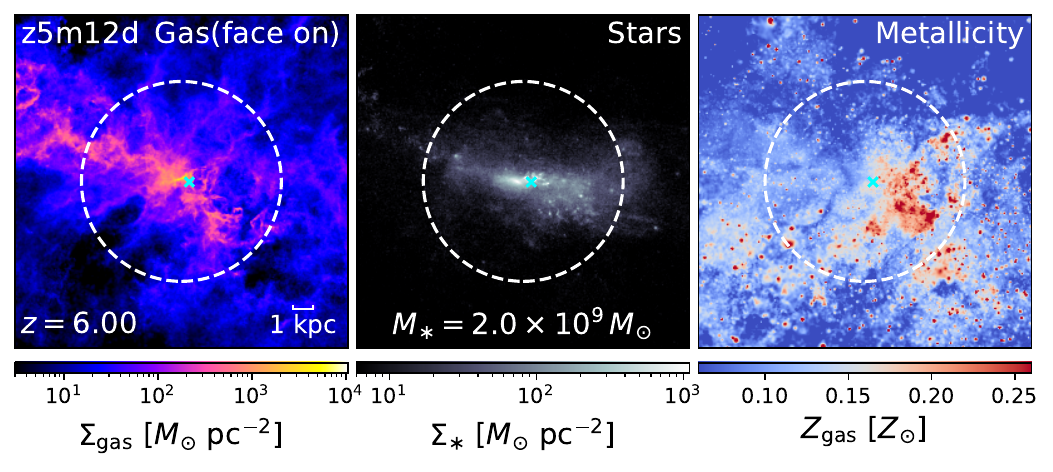}}
 \subfigure{
 \centering
 \includegraphics[align=c,width=0.9\linewidth]{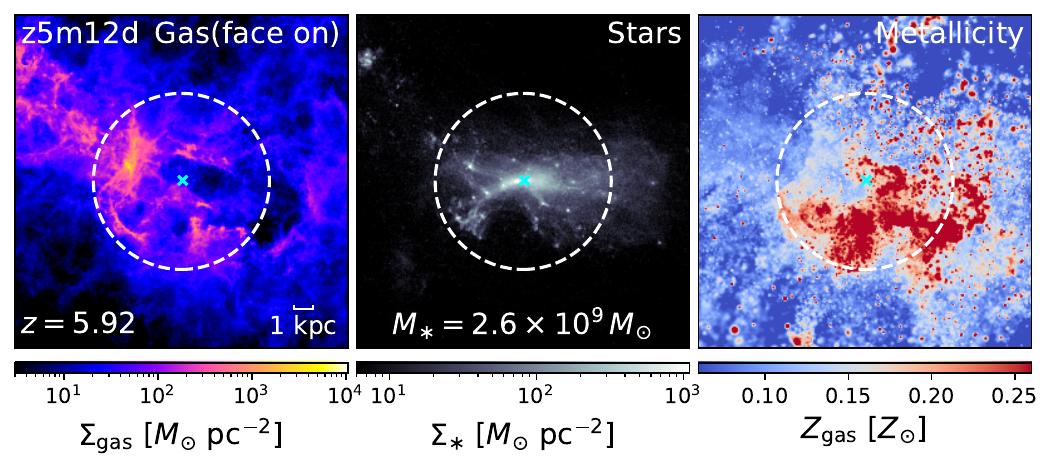}}
\end{minipage}
\begin{minipage}{0.4\linewidth}
 \centering
\subfigure{
 \centering
 \includegraphics[align=c,width=0.92\linewidth]{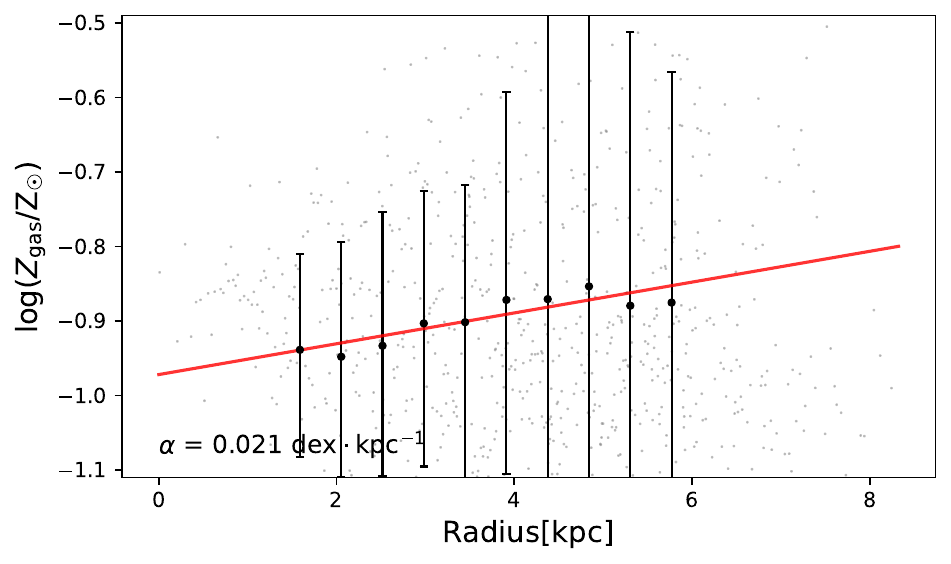}}
 \subfigure{
 \centering
 \includegraphics[align=c,width=0.92\linewidth]{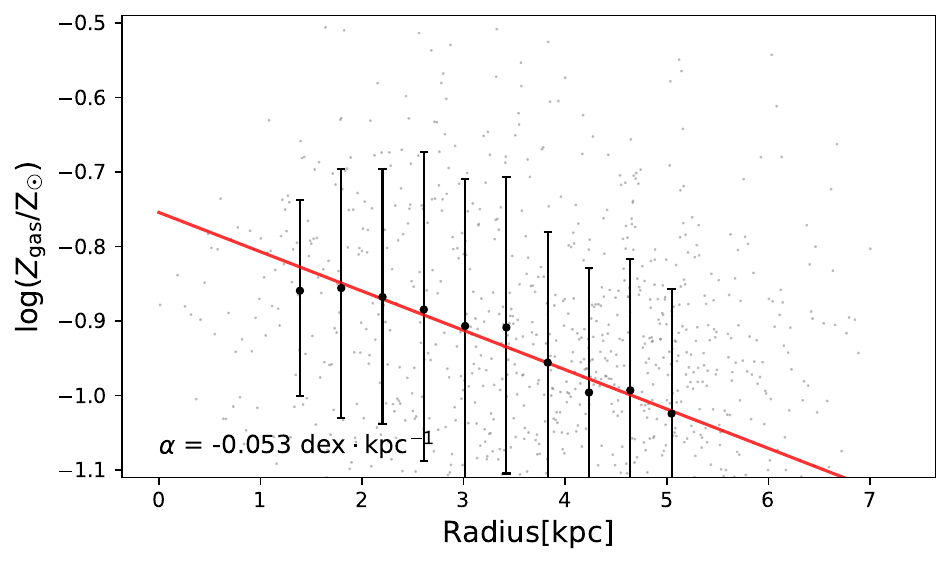}}
 \subfigure{
 \centering
 \includegraphics[align=c,width=0.92\linewidth]{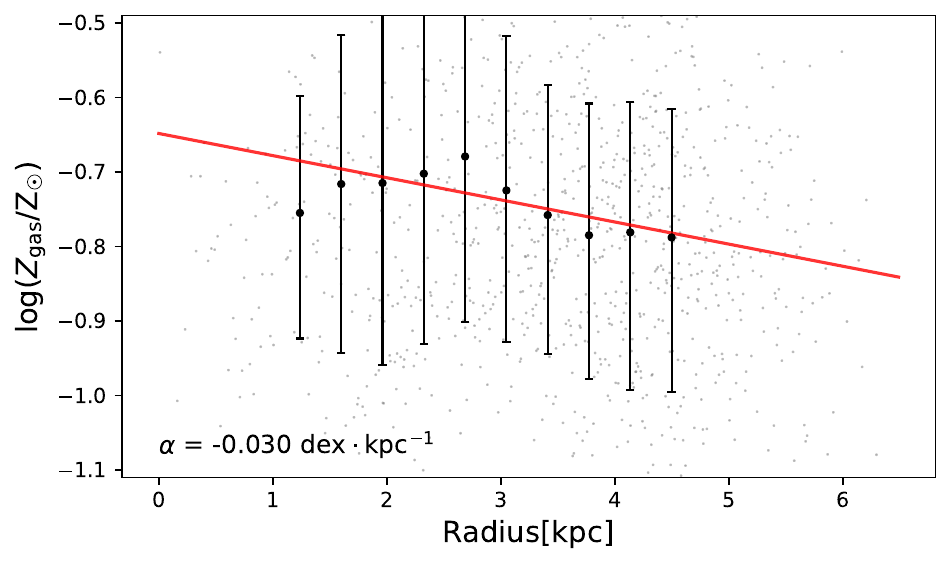}}
\end{minipage}
    \caption{
    \emph{Example galaxy (\texttt{z5m12d}) from our sample}. 
    \emph{Left:} For each sample, the left panel shows the face-on gas density map, the middle panel displays the stellar density map, and the right presents the gas-phase metallicity map. Here the white dashed circle indicates the radius $R_{90}$, within which we measure the metallicity gradient, $5.8\,\mathrm{kpc}$ at $z=6.16$, $5.1\,\mathrm{kpc}$ at $z=6$, $4.6\,\mathrm{kpc}$ at $z=5.92$.
    \emph{Right:} Metallicity gradients of the same sample on the left. The mean values in each bin are shown as black circles, with black error bars representing the corresponding $1-\sigma$ uncertainties. The galaxy exhibits different gradients at different redshift: positive gradient at $z=6.16$, steep negative gradient at $z=6$, flat gradient at $z=5.92$. 
    The evolution from $z=6.16$ to $z=6$ shows a rapid rise in central metallicity (with $z=6.1$ representing an intermediate stage of this enrichment not shown here). This localized central enhancement, driven by localized star formation, produces a very steep negative gradient. By $z=5.92$, feedback becomes effective, driving inside-out enrichment, leading to metal enrichment in the outer regions, which results in a flatter overall gradient.
    }
    \label{fig:gala_def}
\end{figure*}

\subsection{Galaxy Definition and Measurement Method}

In this section, we describe how galaxies are defined for the calculation of metallicity gradients and other physical properties. 
We use Amiga's Halo Finder \citep[\textsc{ahf};][]{Knollmann2009} to identify halos. In this study, we focus on the central galaxy of each snapshot.

We follow the method outlined in \citet{Fitts2017, Ma2017_FIRE_gra} to define the galaxy center as the geometric center of all stellar particles within a shrinking spherical region, iteratively reduced from $20\,\mathrm{kpc}$ to $1\,\mathrm{kpc}$.
A redshift-dependent characteristic radius is then computed based on the empirical size-redshift relation from \citet{Bouwens2004}, using a baseline value of $R = 10\,\mathrm{kpc}$ at $z = 5$ and neglecting stellar mass dependence. Specifically, the initial radius is defined as $R_{\mathrm{initial}} = 60 \cdot(1+z)^{-1}\,\mathrm{kpc}$.
Within this radius, we compute the total stellar mass ($M_\star$) and SFR of the galaxy.
We identify the stellar distribution within this region and determine $R_{90}$, defined as the radius enclosing 90\% of the total mass in young stars with ages $\leq 50\,\mathrm{Myr}$ \citep{Orr2017}. 
This radius is subsequently used as the aperture for computing kinematic quantities and gas-phase metallicity gradients.
To characterize central star formation, we further define a central region as a sphere of radius $0.25\,R_{90}$ centered on the galaxy. The SFR measured within this region is used to quantify centrally concentrated star formation activity.

In the left panel of Fig.~\ref{fig:gala_def}, we present three representative samples from our simulation.
For each galaxy, we align the galaxy's angular momentum with the $z$-axis, such that directions perpendicular and parallel to the $z$-axis correspond to the ``face-on'' and ``edge-on'' views, respectively.
In all panels, a white circle marks the analysis aperture, corresponding to the previously defined $R_{90}$.
The top row of Fig.~\ref{fig:gala_def} shows, from left to right: the face-on gas surface density map, the face-on stellar surface density map, and the face-on gas-phase metallicity map.

To quantify these gradients, we perform a linear fit to the radial metallicity profile over the range $0.25$-$1\,R_{90}$:
\begin{equation}\label{eq:alpha}
\log\left(\frac{Z_{\rm gas}}{Z_\odot}\right) = \alpha R + \beta,
\end{equation}
where $Z_{\rm gas}$ is the gas-phase metallicity and $R$ is the projected radial distance from the galaxy center.
This specific radial range is chosen to improve the robustness of gradient measurements in clumpy and morphologically irregular high-redshift galaxies \citep{Ma2017_FIRE_gra}.
Metallicity gradients in EoR galaxies are often strongly negative; accordingly, we adopt $\alpha = -0.1\,\mathrm{dex\cdot kpc^{-1}}$ as the threshold separating negative gradients ($\alpha < -0.1$) from flat ones ($-0.1 \leq \alpha \leq 0$), and classify gradients with $\alpha > 0$ as positive.

To ensure uniform and adequate radial sampling for our fits, we apply a lower bound of $R_{90}=3\,\mathrm{kpc}$, assigning this floor value to galaxies whose measured $R_{90}$ is below the threshold.
We use only galaxies which $M_\ast\,>\,10^5\mathrm{M_\odot}$.
Furthermore, only gas particles with surface densities above $10\,\mathrm{M_\odot\,pc^{-2}}$ are included, where fragmentation and star formation are expected to occur \citep{Orr2018}.
On the right of Fig.~\ref{fig:gala_def}, we show the gradients of the same three galaxies, which exhibit positive, flat, and steep negative slopes at different redshifts.
As shown in the figure, the metallicity gradient of an individual galaxy can shift from  positive to negative, within $\sim$ two snapshots. This transition timescale is not universal and depends on the strength of feedback; in most cases, it occurs within $\lesssim 50\mathrm{Myr}$.
We also estimate two additional timescales. The free-fall time $\tau_\mathrm{ff} = \sqrt{\frac{3\pi}{32G\rho}}$ \citep{Semenov2017} characterizes the duration of starburst activity, while the mixing time $\tau_\mathrm{mix} = r_\mathrm{gal}/\sigma_\mathrm{turb}$ \citep{Andalman2025} measures the timescale of turbulence-driven gas mixing.
We note that $\tau_\mathrm{mix}$ is intended to approximate mixing following episodes of strong feedback, when gas motions are vigorous enough to impact the metal distribution on galaxy scales; for weaker feedback, mixing is likely more localized and this estimate may not apply.

\begin{figure*}[htbp]
 \centering
 \includegraphics[width=0.91\linewidth]{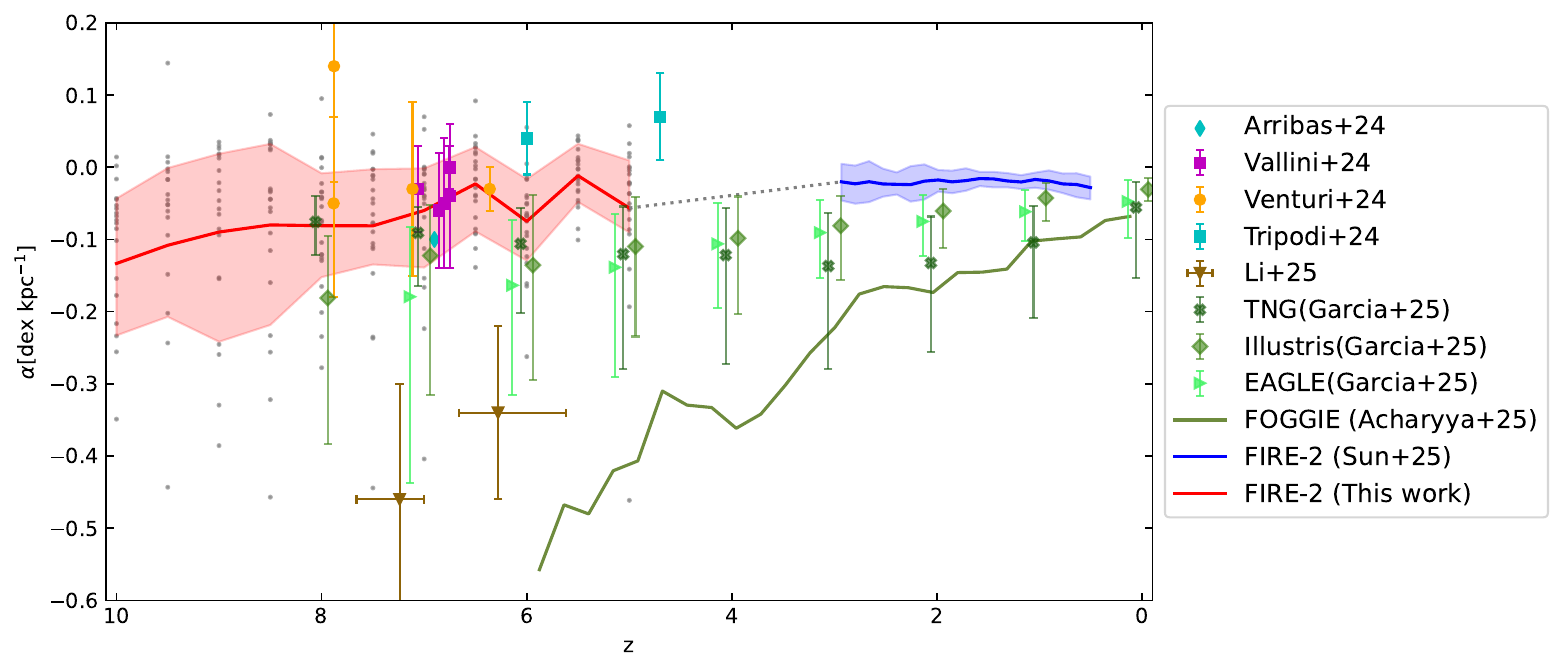}
 \caption{\emph{The cosmic evolution of metallicity gradients.} In general, the FIRE-2 simulations predict that galaxy metallicity gradients flatten from $\sim-0.15~\mathrm{dex\cdot kpc^{-1}}$ in the EoR (i.e. $z>5$) to $\sim$0 (flat radial gradient) during the cosmic noon (i.e. $z\sim2$), broadly consistent with other simulation results \citep{Garcia_EAGLE_TNG, Garcia_SPICE}. The red line and shaded region indicate the median and 1-$\sigma$ spread of our measurements. We also overlay observational results currently available from JWST data. Here the purple points are from \citet{Vallini2024_z7} and \citet{Arribas2024_z7} at $z\sim7$, orange ones from \citet{Venturi2024_z6-8} at $z\sim6-8$, brown ones from \citet{Li2025} and cyan ones from \citet{Tripodi2024_z4-10}. The olive green line is the result of \citet[][FOGGIE]{Acharyya2024_FOGGIE}. The color and symbol scheme is used consistently throughout the following figures. 
Our results indicate that, during the EoR, metallicity gradients become progressively flatter and more stable with decreasing redshift.}
\label{fig:gala_alphatoz}
\end{figure*}

\subsection{Kinematics}

We analyze the kinematic properties of our galaxies following \citet{Yuan2011, Ma2017_FIRE_gra}.
Given that galaxies in the EoR are often irregular, lack well-ordered rotation \citep{Ma2018_FIRE_size}, we adopt a wide slit to better capture their turbulent and irregular velocity fields.
Specifically, a $2\,\mathrm{kpc}$-wide slit is placed along the $z$-axis of each galaxy, sampling gas particles within the region defined by $-R_{90} < y < R_{90}$ and $-1\,\mathrm{kpc} < z < 1\,\mathrm{kpc}$.
All gas particles within this region are used to construct the galaxy's kinematic profile.

For each spatial bin along the $x$-axis, we compute the total line-of-sight momentum and divide by the total gas mass to obtain the mean line-of-sight velocity.
The velocity dispersion, $\sigma$ (or $\sigma_{\mathrm{max}}$), is defined as the maximum line-of-sight velocity dispersion measured across different azimuthal angles within the slit region, capturing the strongest turbulent motions.

We estimate the kinematic properties of each galaxy using two independent methods.
The first approach follows \citet{Leethochawalit2016, Ma2017_FIRE_gra}, in which we compute the peak to peak shear along the slit:
\begin{equation}\label{eq:dvel}
    \Delta v = |\max\{v_{\mathrm{max}}-v_{\mathrm{min}}\}|,
\end{equation}
in three orthogonal directions: line-of-sight ($\Delta v_\mathrm{los}$), radial ($\Delta v_\mathrm{rad}$), and tangential ($\Delta v_\mathrm{tan}$).
We use the deviation to dispersion ratio to simply refer to the ratio of the peak-to-peak maximum velocity shear and velocity dispersion $\Delta v / 2\sigma$.
This quantity characterizes the overall velocity contrast of gas particles in early galaxies, providing a measure of the amplitude of large-scale gas motions relative to random turbulence.
The method offers enhanced flexibility in capturing complex kinematic features-typical of irregular or turbulent systems without assuming symmetry or disk-like structure.
These values are measured as the maximum over all azimuthal angles within the slit.

The second method fits the line-of-sight velocity profile with an arc-tangent function of the form:
\begin{equation}\label{eq:velc}
    v(R)-v_0=v_{\mathrm{c}}\frac{2}{\pi}\arctan\frac{R}{R_{\mathrm{t}}},
\end{equation}
where $v_{\mathrm{c}}$ represents the asymptotic rotation velocity at large radii, and $v_0$ denotes the systemic velocity at the galaxy center \citep[e.g.,][]{Jones2010, Swinbank2012, Stott2014, Leethochawalit2016}.
We restrict $v_\mathrm{c}$ such that $v_\mathrm{c} \lesssim 3\Delta v_\mathrm{los}$ to ensure fit stability.
This approach quantifies the rotational support of the galaxy.
As shown in \citet{Sun2024_z04_3}, we classify the galaxies into three distinct categories based on these fits.
However, since these EoR galaxies often lack coherent rotation, this method can introduce biases when applied to irregular systems.
These two methods complement each other, providing a more comprehensive view of the dynamical state of galaxies.
It should be noted that, based on \citet{Ferreira2022, SunW2024}, disk-like morphology galaxies may exist in EoR, while other studies argue that these structures are prolate spheroids rather than true disks \citep{Vega2024}. In addition, observations have reported kinematic signatures of ordered rotation in EoR galaxies \citep[e.g.][]{Rowland2024}. They use [CII] 158 $\mathrm{\mu m}$ for a massive system $M_\ast\,=\,8^{+4}_{-2}\times10^9\,\mathrm{M_\odot}$, which is substantially more massive than most of our galaxies; therefore, the generally low rotational support in our sample is not unexpected. In this paper, we simply classify galaxy morphology using the rotational support $v_\mathrm{c}/\sigma$; a more detailed discussion of these issues is left for future work.

\section{Results} \label{sec:result}

\subsection{Metallicity gradient versus redshift}\label{subsec:gradtoz}
In Fig.~\ref{fig:gala_alphatoz}, we present the redshift evolution of gas-phase metallicity gradients for all galaxies in our sample across $M_\ast\sim10^5-10^{10}\,\mathrm{M}_{\odot}$. Individual galaxies are shown as gray points, while the red line indicates the average gradient at each redshift. For comparison, blue lines show the results from \citet{Sun2024_z04_3} based on FIRE-2 galaxies at $z \sim 0.44-3$. 
The green points represent simulation results from \citet{Garcia_EAGLE_TNG}, which include data from the TNG, Illustris, and EAGLE projects, covering a mass range of $M_\ast\sim10^{8}-10^{10}\,\mathrm{M}_{\odot}$. The remaining points correspond to observation measurements from \citet{Arribas2024_z7, Vallini2024_z7, Venturi2024_z6-8, Tripodi2024_z4-10, Li2025}, with stellar masses in the range $M_\ast\sim10^{7.5}-10^{11}\,\mathrm{M}_{\odot}$.
This color and symbol scheme is used consistently in the following figures.

In the EoR, FIRE-2 galaxies show negative metallicity gradients with typical slopes of about $-0.1\,\mathrm{dex\cdot kpc^{-1}}$. These values are on average steeper than the mean gradients at cosmic noon, but the scatter is substantially larger, encompassing more extreme cases of both positive and negative gradients.
As galaxies evolve, these gradients systematically flatten, indicating a transition toward a more chemically homogeneous internal structure.
At the same time, scatter in metallicity gradients is also substantially larger at high redshift, reflecting the dynamical instability of early galaxies and allowing for more extreme metallicity gradients, with values ranging from as steep as $-0.5\,\mathrm{dex\cdot kpc^{-1}}$ to as positive as $+0.2\,\mathrm{dex\cdot kpc^{-1}}$.
Over time, however, the scatter decreases steadily, suggesting that galaxies evolve toward more stable and regulated states in both dynamical and chemical properties \citep{Garcia_EAGLE_TNG}. 
This evolution suppresses the occurrence of extreme gradients -- either strongly positive or negative -- at lower redshifts.

These trends are broadly consistent with the results from Illustris and EAGLE at $z > 5$ \citep{Garcia_EAGLE_TNG}, although the metallicity gradients in FIRE-2 tend to be slightly flatter in comparison. This is likely due to differences in metal mixing: EAGLE locks metals to individual gas particles, such that metals are not exchanged between particles \citep{Aguirre2005}, and Illustris lacks explicit modeling of unresolved turbulent mixing \citep{Escala2018, Semenov2024}. These differences can lead to less efficient metal redistribution and thus steeper gradients in those simulations.
FOGGIE \citep{Acharyya2024_FOGGIE} results exhibit strongly negative gradients. As mentioned in that paper, the relatively weak supernova feedback and centrally concentrated star formation tend to retain metals in the central regions, thereby producing systematically steeper negative gradients.

The metallicity gradients predicted by FIRE-2 are broadly consistent with recent high-redshift measurements \citep{Venturi2024_z6-8, Vallini2024_z7}: despite the substantial observational uncertainties, most simulated values fall within the reported $1-\sigma$ ranges.
FIRE-2 also produces a subset of galaxies with steep positive or negative gradients, in line with the diversity inferred from current observations \citep{Venturi2024_z6-8, Li2025}.
Taken together, this suggests that the FIRE-2 model captures the main drivers of metal enrichment and redistribution in early galaxies.



\subsection{Metallicity gradient versus stellar mass}

\begin{figure}[htbp]
  \hspace{-0.2cm}
 \centering
 \includegraphics[align=c,width=\linewidth]{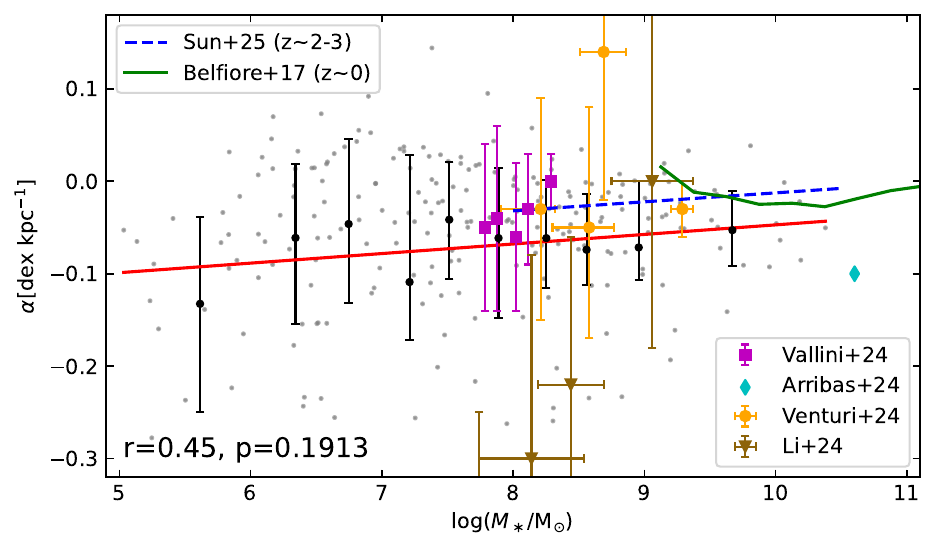}
 \caption{\emph{Metallicity gradient versus stellar mass.}
The red line indicates the best-fit linear relation.
We find a positive correlation between gas-phase metallicity gradient and stellar mass, consistent with recent observational results \citep{Vallini2024_z7,Venturi2024_z6-8,Li2025}.
For comparison, we also show results at other redshifts \citep{Sun2024_z04_3, Belfiore2017}.
}
\label{fig:gala_alphatoM}
\end{figure}

\begin{figure}[htbp]
  \hspace{-0.2cm}
 \centering
 \includegraphics[align=c,width=\linewidth]{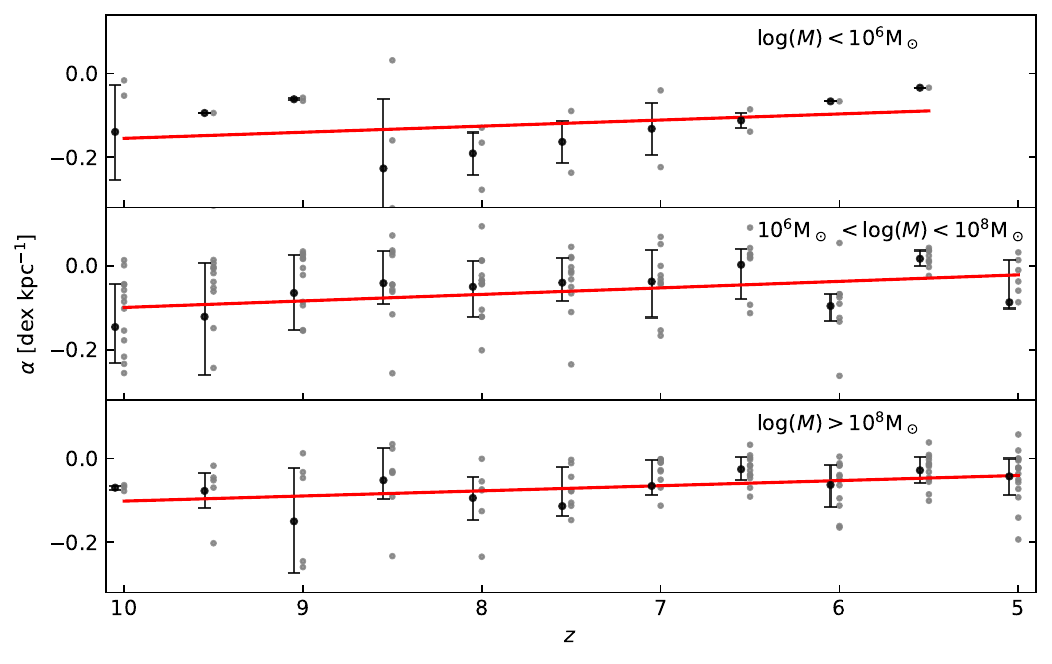}
 \caption{\emph{Metallicity gradient versus redshift at different stellar masses.}
Galaxies are divided into three stellar mass bins: $M_\star < 10^6\,\mathrm{M}_\odot$, $10^6\,\mathrm{M}_\odot < M_\star < 10^8\,\mathrm{M}_\odot$, and $M\star > 10^8\,\mathrm{M}_\odot$.
This binning allows us to examine how the redshift evolution of gas-phase metallicity gradients depends on galaxy stellar mass.
Massive galaxies tend to show a more gradual evolution in their metallicity gradients with redshift, in contrast to low-mass galaxies, which display more pronounced and rapid changes over time.}
\label{fig:a2m_z}
\end{figure}

In Fig.~\ref{fig:gala_alphatoM}, we present the relationship between gas-phase metallicity gradient and stellar mass.
A clear trend emerges: more massive galaxies tend to exhibit flatter (less negative) metallicity gradients.
This behavior is broadly consistent with observational results at cosmic noon, where a positive correlation between gradient and stellar mass is observed below $M_\star \sim 10^{10}\,\mathrm{M}_\odot$ \citep{Stott2014, Ma2017_FIRE_gra, Sun2024_z04_3}.
Low-mass systems tend to show steeper gradients and larger scatter, reflecting their chaotic internal structures.
In contrast, the scatter in metallicity gradients decreases significantly toward the high-mass end, suggesting that more massive galaxies develop dynamically stable configurations capable of maintaining relatively uniform chemical distributions, even in the EoR.
By contrast, in the modern Universe the trend appears to be governed by a different mechanism: metallicity gradients become more negative with increasing stellar mass below $M_\star < 10^{10.5}\,\mathrm{M}_\odot$, but flatten again toward higher masses \citep{Hemler2021, Sharda2021b}.

Given the limited number of high-redshift measurements and the large scatter in current data, a precise universal relation cannot yet be established. 
This supports the interpretation that stellar mass is a primary driver of internal chemical structure, and that this connection is already in place at early times, implying a rapid emergence of the mass--metallicity distribution link during galaxy formation \citep{Baker2023, Menguiano2024}.

Fig.~\ref{fig:a2m_z} further illustrates the redshift evolution of metallicity gradients as a function of stellar mass.
Galaxies are grouped into three mass bins: $M_\star < 10^6\,\mathrm{M}_\odot$ (top), $10^6\,\mathrm{M}_\odot < M_\star < 10^8\,\mathrm{M}_\odot$ (middle), and $M_\star > 10^8\,\mathrm{M}_\odot$ (bottom).
Here, a strong mass dependence is evident: low-mass galaxies exhibit substantial redshift evolution in their gradients, while high-mass systems maintain relatively constant gradients over time.

This mass-dependent behavior is likely linked to differences in regulatory mechanisms. In high-mass galaxies, as their physical size increases, internal feedback processes -- particularly gas flows -- tend to become more effective.
Massive systems also accrete more gas \citep{Yates2021, Sharda2021b}, and during the EoR such enhanced accretion can drive more vigorous gas inflow--outflow cycling and radial gas exchange.
These gas flows facilitate the redistribution of metals, resulting in more stable and flatter metallicity gradients, along with smoother and more coherent chemical evolution \citep{Ma2017_FIRE_gra, Graf2024}. 
In contrast, low-mass galaxies experience more bursty star formation, and their feedback processes often fail to mix metals efficiently on short timescales. This leads to steep negative metallicity gradients as metal-enriched gas remains concentrated near its origin, as seen in Fig.~\ref{fig:gala_alphatoM}.

\subsection{Metallicity gradient versus SFR}

\begin{figure*}[htbp]
  \hspace{-0.2cm}
\subfigure{
 \label{fig:gala_alphatoSFR}
 \centering
 \includegraphics[align=c,width=0.47\linewidth]{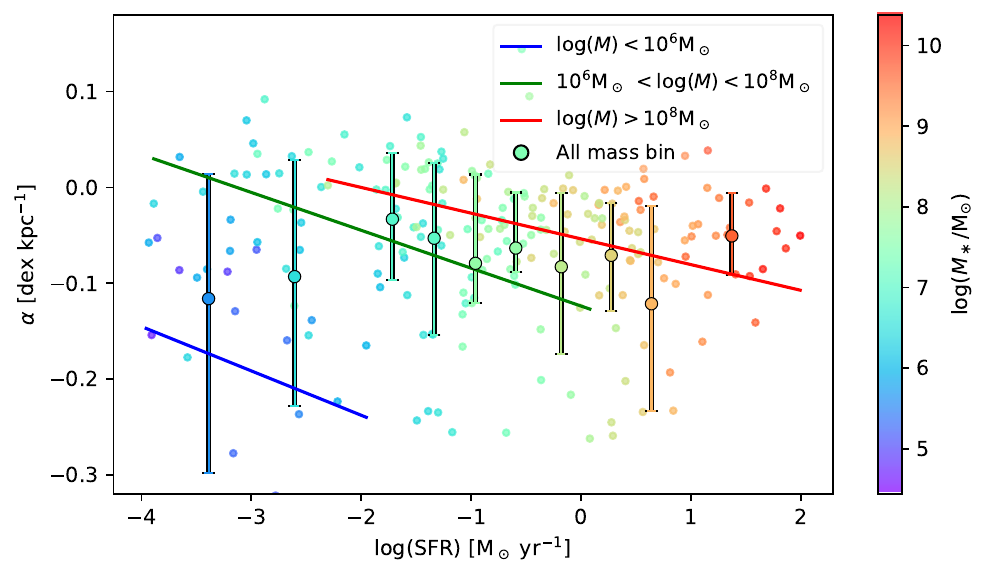}}
  \hspace{0.2cm}
\subfigure{
 \label{fig:gala_alphatosSFR}
 \centering
 \includegraphics[align=c,width=0.47\linewidth]{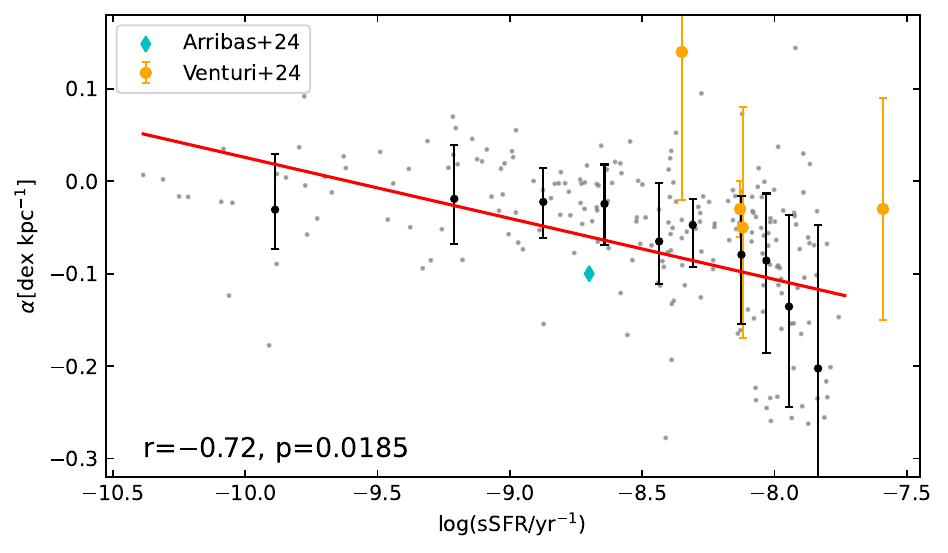}}
 \caption{
 \emph{Left: Metallicity gradient versus SFR.}
 SFR is measured as the young stars over the past $50\,\mathrm{Myr}$.
 The color of each point shows the stellar mass of galaxy.
 The error-bar shows the means and $1-\sigma$ of all samples. We also divide these galaxies into different stellar mass bins as linear fits in different color $M_\star <10^6\,\mathrm{M_\odot}$, $10^6\,\mathrm{M_\odot} <M_\star <10^8\,\mathrm{M_\odot}$ and $M_\star >10^8\,\mathrm{M_\odot}$.
 Across different mass bins, the metallicity gradient shows a similar trend, becoming steeper as galaxies become more actively star-forming.
 \emph{Right: Metallicity gradient versus sSFR.}
 Here the red line shows the linear fit.
 The metallicity gradients exhibit a strong negative correlation with sSFR, which is completely opposite to the trend during the cosmic noon epoch \citep[see Fig.~4 in][]{Sun2024_z04_3}.
 }
\label{fig:gala_alphatosSSFR}
\end{figure*}


In Fig.~\ref{fig:gala_alphatosSSFR}, we examine the relationship between gas-phase metallicity gradients with both SFR and sSFR, measured over the past $50\,\mathrm{Myr}$.

We find no correlation between the metallicity gradient and SFR when considering the entire galaxy sample. The left panel of Fig.~\ref{fig:gala_alphatosSSFR} suggests that variations in total star formation activity have a limited effect on the overall gradient.
However, this is not the case when stellar mass is taken into account. As shown in the different mass bins of Fig.~\ref{fig:gala_alphatosSSFR}, we still examine the relationship between metallicity gradients and SFR across three stellar mass bins: $M_\star < 10^6\,\mathrm{M}_\odot$, $10^6\,\mathrm{M}_\odot < M_\star < 10^8\,\mathrm{M}_\odot$, and $M_\star > 10^8\,\mathrm{M}_\odot$.
Across all mass bins, metallicity gradients show a clear negative correlation with SFR, suggesting that ongoing star formation plays a significant role in regulating the internal chemical structure of galaxies.
In the highest mass bin, the correlation between metallicity gradient and SFR becomes weaker. 
Such regulation suppresses the development of steep gradients, even in the presence of vigorous star formation.

Metallicity gradients exhibit a similar negative correlation with sSFR, a trend similar with \citet{Venturi2024_z6-8}.
This contrasts with the low-redshift trend reported by \citet{Stott2014, Ma2017_FIRE_gra, Sun2024_z04_3}, where metallicity gradients become flatter with increasing sSFR.
In high-redshift sample ($z \gtrsim 5$), the opposite behavior is observed: higher sSFR corresponds to steeper negative gradients.
This reversal likely reflects differences in the relative timescales of star formation and metal mixing.
At high redshift, star formation is more bursty and localized, often occurring in dense clumps where the free-fall timescale can be as short as $\tau_{\mathrm{ff}}\sim10\,\mathrm{Myr}$, the metals produced by these stars are subsequently released during supernovae, leading to rapid metal enrichment in compact regions.
However, the metal mixing timescale for these galaxies $\tau_{\mathrm{mix}}\sim10\,\mathrm{Myr}$ is typically too long to homogenize the newly enriched metals, leading to significant chemical inhomogeneities.
As a result, the presence of numerous metal-enriched clumps within these galaxies are frequently observed, making them show strong azimuthal patchiness.

By comparing the relationships between metallicity gradients and both SFR and sSFR, we find that the interplay between star formation and internal chemical structure is strongly modulated by stellar mass.
While the strength of stellar feedback (as traced by SFR) contributes to shaping the metallicity gradient, the relative efficiency of feedback per unit stellar mass (sSFR) plays a more direct regulatory role by influencing gas flows within galaxies.

\subsection{Metallicity gradient versus kinematic parameters}

\begin{figure*}[htbp]
  \hspace{-0.2cm}
\subfigure{
 \label{fig:gala_alphatos}
 \centering
 \includegraphics[align=c,width=0.47\linewidth]{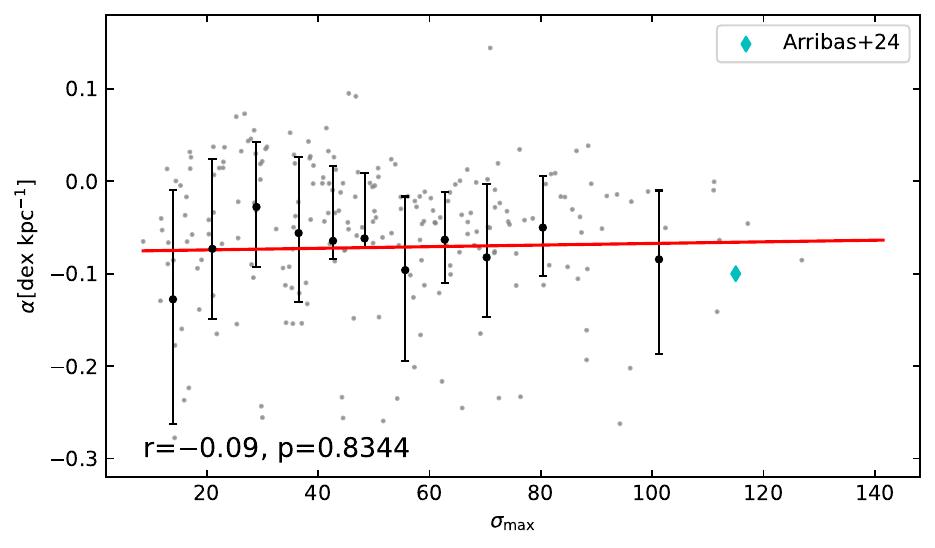}}
  \hspace{0.2cm}
\subfigure{
 \label{fig:gala_alphatovs}
 \centering
 \includegraphics[align=c,width=0.47\linewidth]{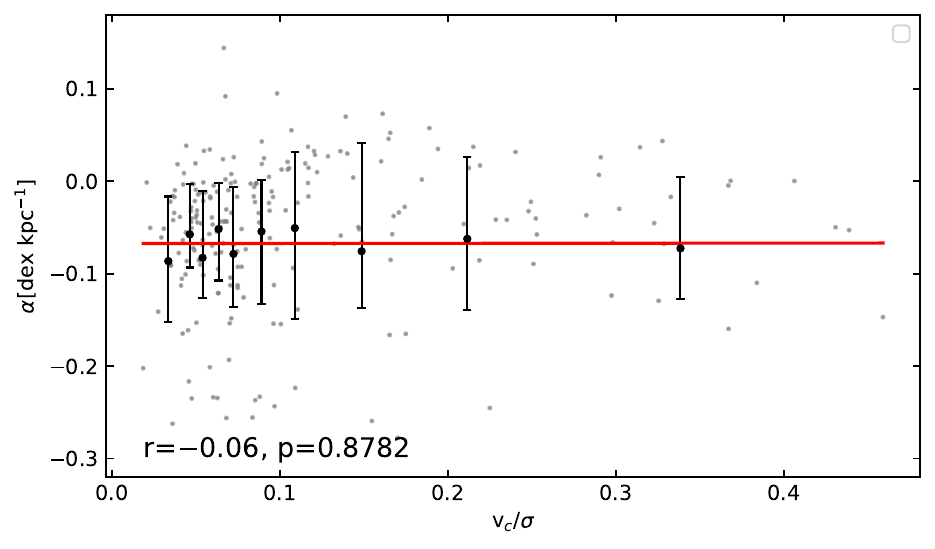}}
 \caption{\emph{Metallicity gradient versus velocity dispersion $\sigma$ and rotational support $v_\mathrm{c}/\sigma$ for all gas particles.} These EoR galaxies predominantly exhibit irregular morphologies.
 The red line represents the linear fit of our samples. 
 The metallicity gradient exhibits basically no correlation with velocity dispersion and $v_\mathrm{c}/\sigma$.
 }
\label{fig:gala_alphatos-vs}
\end{figure*}

\begin{figure}[htbp]
  \hspace{-0.2cm}
\subfigure{
 \centering
 \includegraphics[align=c,width=\linewidth]{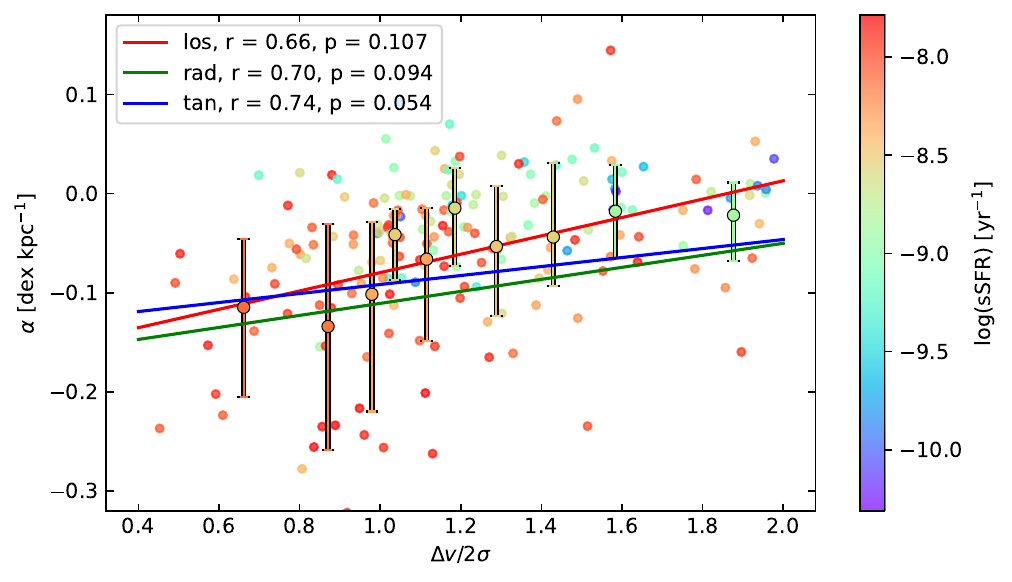}}
 \caption{ 
 \emph{Metallicity gradient versus velocity deviation to dispersion ratio $\Delta v/2\sigma$ for all gas particles in three directions.} The color of each point represents the galaxy's sSFR. 
 The big dots and error bars represent the median and $1-\sigma$ uncertainty values measured in each $\Delta v_\mathrm{los}/2\sigma$ bin, colored-coded in the average sSFR.
 The metallicity gradient becomes steeper (i.e., more negative) with decreasing $\Delta v_\mathrm{los}/2\sigma$, suggesting that star-forming gas particles are increasingly concentrated toward the galaxy center, further supported by the elevated sSFR.
 A similar trend is found for the radial and tangential directions.
 }
 \label{fig:gala_alphatodvs}
  \hspace{0.2cm}
\end{figure}

In this section, we examine the correlations between metallicity gradients and key kinematic indicators: gas velocity dispersion ($\sigma$), the degree of rotational support ($v_\mathrm{c}/\sigma$), and the deviation to dispersion ratio ($\Delta v/2\sigma$), measured within $R_{90}$ for all gas particles.

Following \citet{Kassin2012, Girard2020}, galaxy with $v_\mathrm{c}/\sigma>1$ is classified as ``rotationally supported'', all EoR galaxies are expected to lack well-ordered disk structures and instead exhibit irregular or clumpy morphologies in this study. Consequently, strong correlations between structural kinematics and metallicity gradients may not yet be established. This is reflected in the left panel of Fig.~\ref{fig:gala_alphatos-vs}, where metallicity gradients show no clear linear correlation with velocity dispersion.
While hints of non-monotonic behavior (e.g. rise-fall-rise trends) may be present, the current sample size is insufficient to draw firm conclusions.

Similarly, the right panel of Fig.~\ref{fig:gala_alphatos-vs} shows only a very weak correlation between metallicity gradients and $v_\mathrm{c}/\sigma$.
Moreover, the values of $v_\mathrm{c}/\sigma$ are generally low, with the majority clustered at $\lesssim 0.2$, which further limits the statistical significance of any observed trend.
This reinforces the interpretation that coherent rotational support is not yet established in almost all EoR galaxies.
We note, however, that this refers to the lack of a clear linear correlation within the limited $v_\mathrm{c}/\sigma$ dynamic range; it does not rule out the presence of non-linear or higher-order dependencies.

In contrast, Fig.~\ref{fig:gala_alphatodvs} suggests a positive correlation between metallicity gradient and $\Delta v_\mathrm{los}/2\sigma$: as $\Delta v_\mathrm{los}/2\sigma$ increases, gradients become progressively flatter. The same trend holds for the radial and tangential directions.
This indicates that high-velocity gas flows play a critical role in regulating the spatial distribution of metals in irregular galaxies.
Such flows can flatten gradients through a combination of metal-rich outflows, inflows of low-metallicity gas that dilute central enrichment, or enhanced turbulent mixing across large spatial scales.
As previously discussed, intense starbursts rapidly enrich the dense central regions of galaxies, producing steep negative gradients. 
Feedback-driven winds then redistribute metals more uniformly throughout the ISM, flattening the gradient and promoting chemical homogeneity.
These gas flows can also suppress star formation by reducing density and disrupting the formation of cold, dense gas clumps.

Taken together, these findings suggest that fitting high-redshift galaxies with a standard $\arctan$ rotation curve -- typically applied to well-ordered disk systems -- may be inappropriate.
Such models assume symmetric, rotationally supported structures that are often absent for EoR galaxies. 
In contrast, the deviation to dispersion ratio $\Delta v/2\sigma$ provides a more flexible and physically meaningful tracer of the gas kinematics that regulate metal transport, especially in galaxies that have not yet developed stable disk-like structures.

\subsection{Localized star formation}

\begin{figure}[htbp]
  \hspace{-0.2cm}
\subfigure{
 \centering
 \includegraphics[align=c,width=\linewidth]{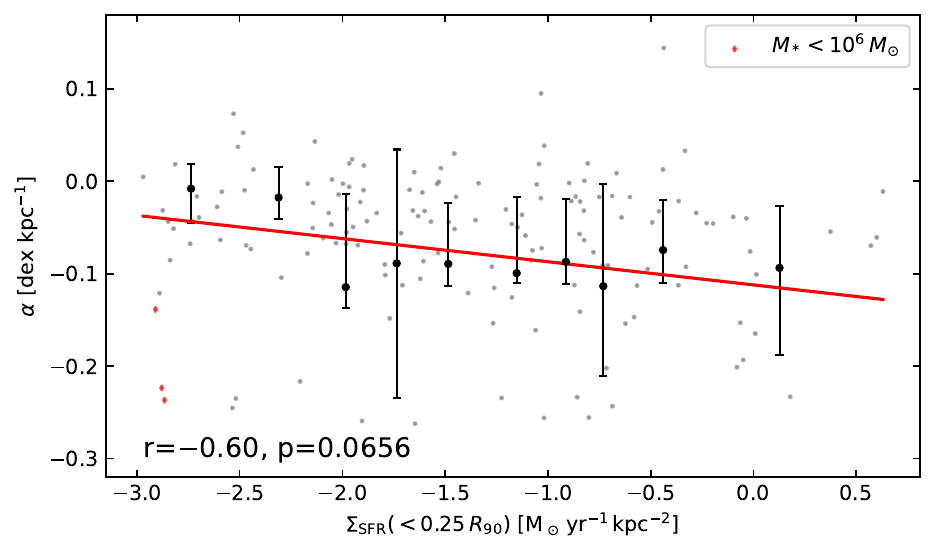}}
 \caption{ 
 \emph{Metallicity gradient versus the SFR surface density in the central region $r<0.25\,R_{90}$.} A higher central SFR density indicates more intense localized star formation, significantly enriching the galaxy's central regions with newly produced metals and thus producing a steeper metallicity gradient.}
 \label{fig:gala_alphatocenSFR}
  \hspace{0.2cm}
\end{figure}

In this section, we examine how gas-phase metallicity gradients relate to centrally concentrated star formation. We quantify the localized star-formation activity using the central SFR surface density, $\Sigma_{\mathrm{SFR}}(<0.25\,R_{90})$, where the central region is defined as  $r < 0.25\,R_{90}$. Our fits include galaxies with $\Sigma_{\mathrm{SFR}}(<0.25\,R_{90}) \,>\,-3.0\,\mathrm{M_\odot\,\mathrm{yr}^{-1}\,\mathrm{kpc}^{-2}}$ and $M_\ast\,>\,10^6\mathrm{M_\odot}$.

Fig.~\ref{fig:gala_alphatocenSFR} reveals a negative correlation, similar to the previously discussed sSFR trend: galaxies with higher central star formation activity tend to exhibit steeper negative metallicity gradients. This relationship likely arises from localized enrichment processes, where intense, centrally concentrated starbursts rapidly inject metals into the galaxy's core, creating pronounced radial contrasts in metallicity.

For these EoR galaxies, redistributing metals over galactic scales requires substantial gas flows and turbulence.
These mixing timescales ($\tau_\mathrm{mix}\sim10\,\mathrm{Myr}$) are comparable to the bursty timescales of star formation episodes in young galaxies. 
As a result, gas motions are unable to effectively homogenize metals across the galaxy within short intervals, significantly limiting their spatial transport over large distances \citep[see also in massive galaxies:][]{Bassini20231}. 
Metals remain confined near their production sites, creating numerous metal-rich clumps.
Since our observational centers are intentionally defined at the highest-density cores, regions typically dominated by intense starburst activity, the steep negative metallicity gradients produced by localized enrichment become especially pronounced and readily measurable.
This picture is also consistent with \citet{SunGC2026}, who showed that in these high-redshift bursty galaxies, turbulence can generate localized pockets of dense gas where star formation proceeds with high efficiency, while the galaxy-averaged star formation efficiency (SFE) remains relatively low.

As shown in the first two snapshots of Fig.~\ref{fig:gala_def}, bursty star formation episodes in the central regions drive a rapid increase in metallicity. 
However, mixing on galactic scales is relatively inefficient at this redshift, which limits metal redistribution and allows metals to continuously accumulate in the center, developing steep negative gradients. 
Feedback then becomes effective in the subsequent phase.
Whether through star formation in the outskirts that releases new metals, or through powerful gas flows that redistribute existing metals, the previously steep gradients are smoothed out once again.

Thus, these EoR galaxies differ from their cosmic noon counterparts, where stronger feedback processes drive large-scale gas motions and turbulence that smooth metallicity gradients across galactic scales.
At lower redshift ($z<1$), once stable rotating disks are in place, secular processes such as inside-out growth, continuous central star formation, and bar/spiral torque-driven radial inflows-maintain the central build-up of metals and produce smooth, negative metallicity gradients.

\section{Conclusions} \label{sec:conclusion}

In this paper, we utilize the high-redshift suite of the FIRE-2 cosmological hydrodynamic simulations to investigate the evolution of gas-phase metallicity gradients in galaxies from $z\sim10$ down to $z\sim5$, spanning the EoR.
Our analysis is based on 22 simulated galaxies covering a wide range of stellar masses.

By quantifying the relationship between gradients and galaxy properties such as stellar mass, star forming, and gas velocity metrics, we provide deeper insights into the physical mechanisms responsible for shaping the chemical structures of EoR galaxies.

\begin{itemize}
    \item In the EoR, galaxies show more extreme metallicity gradients, often much steeper than those seen at cosmic noon. 
    As galaxies evolve, these gradients tend to flatten and the scatter decreases, consistent with a progressive transition toward greater dynamical stability that reduces the prevalence of strongly positive or negative gradients at later times.
    \item Metallicity gradients correlate positively with stellar mass over $M_\ast\sim10^5-10^{10}\,\mathrm{M}_\odot$. 
    Massive galaxies behave more stably, tend to have flatter average gradients and smaller scatter, and remain relatively constant over time. 
    In contrast, low-mass galaxies show stronger redshift evolution.
    \item When galaxies are compared at fixed stellar mass, higher SFR is associated with more negative metallicity gradients (equivalently, the same trend is seen with sSFR), even though this relation is washed out in the full, unbinned sample.
    These starbursts likely operate differently from those at cosmic noon: newly produced metals are not mixed rapidly and instead remain near their formation sites, especially in the central regions, leading to steep metallicity gradients.
    \item 
    The link between metallicity gradients and kinematics in EoR galaxies is not well captured by standard rotation/dispersion measures. 
    In these irregular systems, $\sigma_\mathrm{max}$ and $v_\mathrm{c}/\sigma$ are not robust descriptors and show no clear association with the gradients. 
    A clearer signal appears when using $\Delta v/2\sigma$, which correlates positively with the gradients and largely tracks sSFR. 
    A plausible interpretation is an evolutionary sequence: central starbursts rapidly enrich the inner regions, producing steep negative gradients, while subsequent feedback injects turbulence, enhances mixing, and redistributes metals, progressively flattening the gradients.
\end{itemize}

EoR galaxies differ markedly from their cosmic-noon counterparts in both structural maturity and the dominant internal processes.
In these early systems, metal distributions are shaped by bursty, spatially localized star formation.
Starbursts preferentially occur in dense central regions, depositing freshly produced metals into the surrounding ISM, building metal-rich clumps and rapidly enriching the galactic core.
Crucially, the large-scale redistribution timescale, $\tau_{\rm mix}$, is comparable to the duration of individual burst episodes.
Mixing therefore cannot act efficiently on $\sim10\,\mathrm{Myr}$ timescales, and enriched gas remains close to its production sites. 

The result is a chemically enriched core contrasted with relatively pristine outskirts, producing steep radial metallicity gradients and strong spatial inhomogeneities. Such gradients are often short-lived, since subsequent feedback-driven turbulence can mix and redistribute metals, but the cycle of localized enrichment with weak internal regulation is a hallmark of early galaxy evolution—distinct from the more self-regulated, chemically homogenized systems at later times.

Building on these results, we propose the following evolutionary picture. 
During the EoR, star formation is bursty and spatially localized, while the metal-mixing timescale remains relatively long; newly enriched gas therefore stays near its production sites, producing steep negative metallicity gradients. 
By cosmic noon, sustained and strong feedback drives efficient, large-scale mixing and metal transport, rapidly redistributing metals across galaxies and thereby flattening—sometimes even inverting—the gradients. 
At later times, as galaxies settle into rotationally supported disks, feedback becomes less effective at globally stirring the ISM, allowing metals to accumulate more steadily in the central regions and leading to the re-establishment and long-term stabilization of negative gradients.
In future work, we may further combine radiative transfer modeling with these samples to explore the metal distribution of these galaxies in a way that more closely resembles observational methods.


\begin{acknowledgments}
We thank the anonymous referee for a constructive report that helps improve the clarity of this paper.
This work is supported by the National Key R\&D Program of China No.2025YFF0510603, the National Natural Science Foundation of China (grant 12373009), the CAS Project for Young Scientists in Basic Research Grant No. YSBR-062, the China Manned Space Program with grant no. CMS-CSST-2025-A06, and the Fundamental Research Funds for the Central Universities. XW acknowledges the support by the Xiaomi Young Talents Program, and the work carried out, in part, at the Swinburne University of Technology, sponsored by the ACAMAR visiting fellowship.
Part of the numerical calculations in this study were carried out on the ORISE Supercomputer (grant DFZX202510).
LCH was supported by the National Science Foundation of China (12233001) and the China Manned Space Program (CMS-CSST-2025-A09).
AW received support from NSF, via CAREER award AST-2045928 and grant AST-2107772, and HST grant GO-16273 from STScI.
\end{acknowledgments}

\appendix

\begin{figure}[htbp]
  \hspace{-0.2cm}
\subfigure{
 \centering
 \includegraphics[align=c,width=\linewidth]{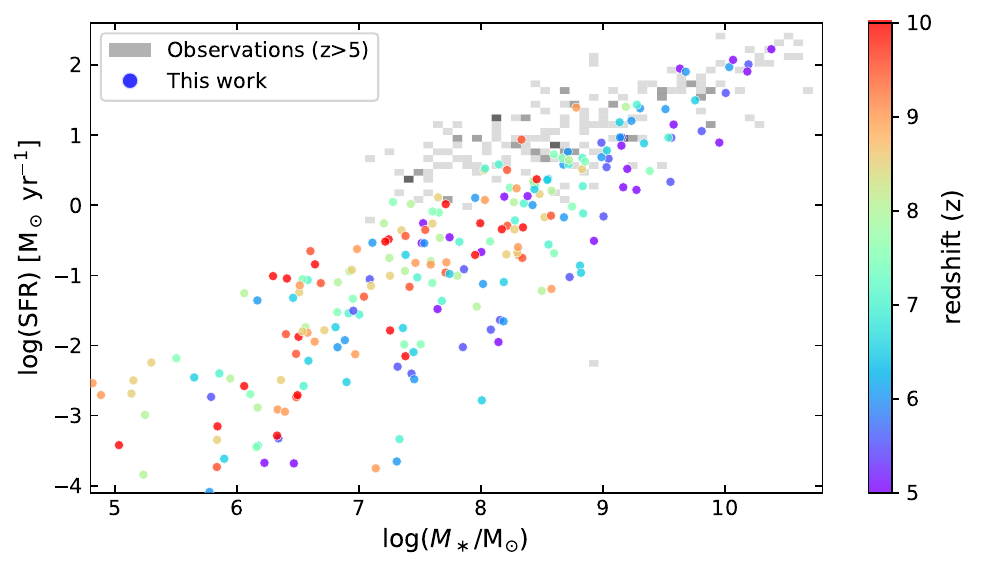}}
 \caption{ 
 \emph{SFR versus stellar mass.} The color of each point shows the redshift of all simulation galaxies in this work. The grey-scale 2D histogram represents the number density of galaxies at $z>5$ observed with JWST.
 }
 \label{fig:gala_SFR2Mass}
  \hspace{0.2cm}
\end{figure}

\begin{figure}[htbp]
  \centering
  \subfigure{
    \label{fig:gala_z2s}
\includegraphics[align=c,width=\linewidth]{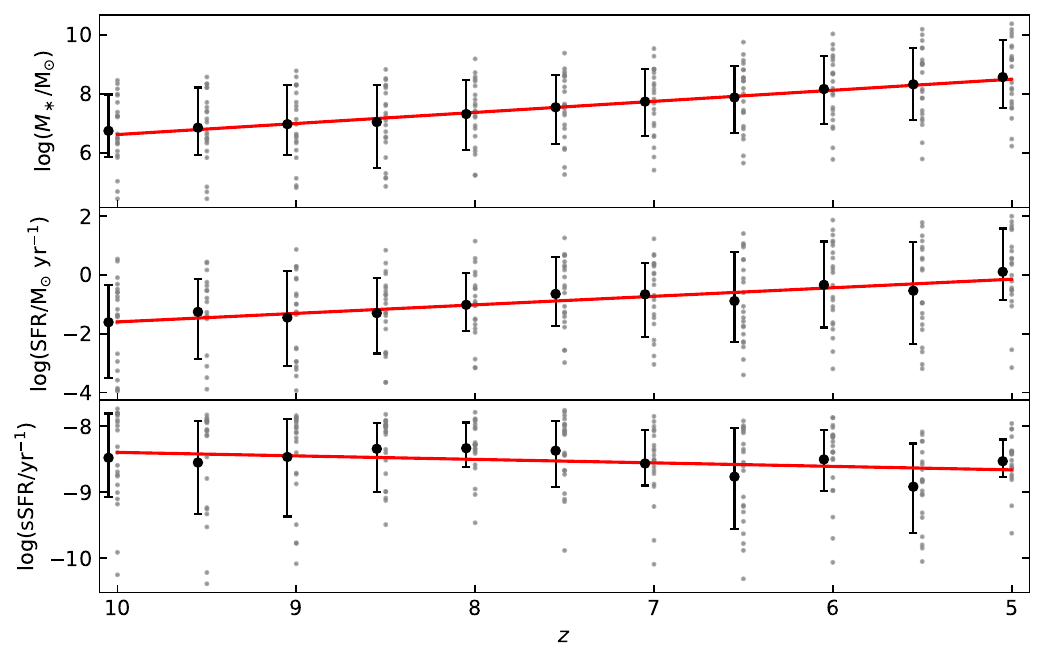}
  }
  \vspace{0.3cm}
  \subfigure{
    \label{fig:gala_z2v}
\includegraphics[align=c,width=\linewidth]{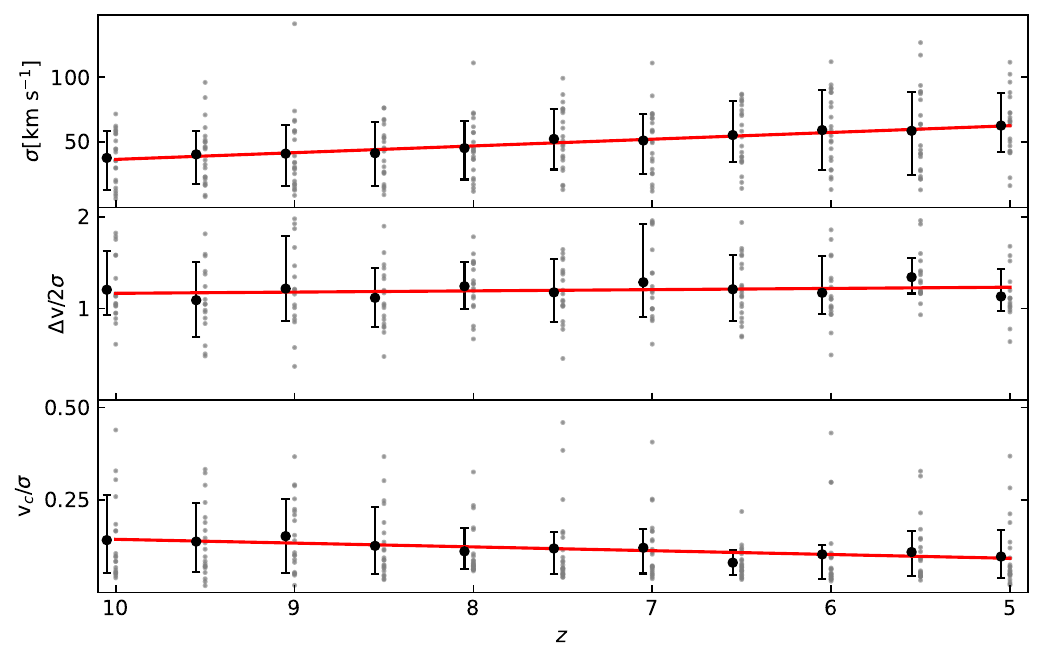}
  }
  \caption{\emph{Top:} stellar mass $M_\star$, SFR and sSFR versus redshift. 
  \emph{Bottom:} velocity dispersion $\sigma_{\mathrm{max}}$, degree of rotational support $v_\mathrm{c}/\sigma$ and deviation to dispersion ratio $\Delta v_\mathrm{los}/2\sigma$ for all gas particles versus redshift.
  Stellar mass, SFR, and $\sigma_{\mathrm{max}}$ exhibit strong positive evolution with redshift, whereas $v_\mathrm{c}/\sigma$ shows a pronounced negative redshift evolution. 
  In contrast, $\Delta v_\mathrm{los}/2\sigma$ exhibits no significant evolution with redshift.}
  \label{fig:gala_ztopros}
\end{figure}

\section{Galaxy properties evolution}\label{app:evo_z}
In this appendix, we investigate the relationship between SFR and stellar mass in Fig.\ref{fig:gala_SFR2Mass}.
We also show the observation results from \citet{Nakajima2023, Gottumukkala2023, Harikane2023, Sarkar2025, Scholtz2025, Hu2025}. 
Our results are broadly consistent with most observations, though some observed galaxies appear significantly more active than predicted by FIRE-2. This discrepancy is likely driven by upscatter effects: short-lived bursts of star formation can temporarily elevate the observed SFR \citep[e.g.][]{SunG2023, Shen2023, Kravtsov2024}, whereas the SFR metric used in our analysis, based on a $50\,\mathrm{Myr}$ average, cannot capture such transient enhancements.

The redshift evolution of several key galaxy properties is presented in Fig.~\ref{fig:gala_ztopros}. 
Stellar mass increases significantly over time, reflecting the ongoing assembly of galaxies. The SFR also rises with decreasing redshift, but at a slower pace than the mass growth, resulting in a steady decline in sSFR.
Velocity dispersion $\sigma$ gradually increases with time, suggesting that while galaxies grow in size, they have not yet developed well-ordered structures. The rise in dispersion may be attributed to the larger gas reservoirs and more energetic feedback processes in more massive galaxies.
Similarly, $\Delta v/2\sigma$ shows no significant evolution with redshift, implying that the value of this ratio is not strongly dependent on redshift but may be influenced by other factors, such as the number of gas particles sampled or the velocity of gas particles.
Likewise, $v_\mathrm{c}/\sigma$ remains consistently low throughout the redshift range, indicating a persistent lack of rotational support in these dynamically unsettled high-redshift galaxies.

\section{The evolution of each galaxy}

\begin{figure}[htbp]
  \hspace{-0.2cm}
\subfigure{
 \centering
 \includegraphics[align=c,width=\linewidth]{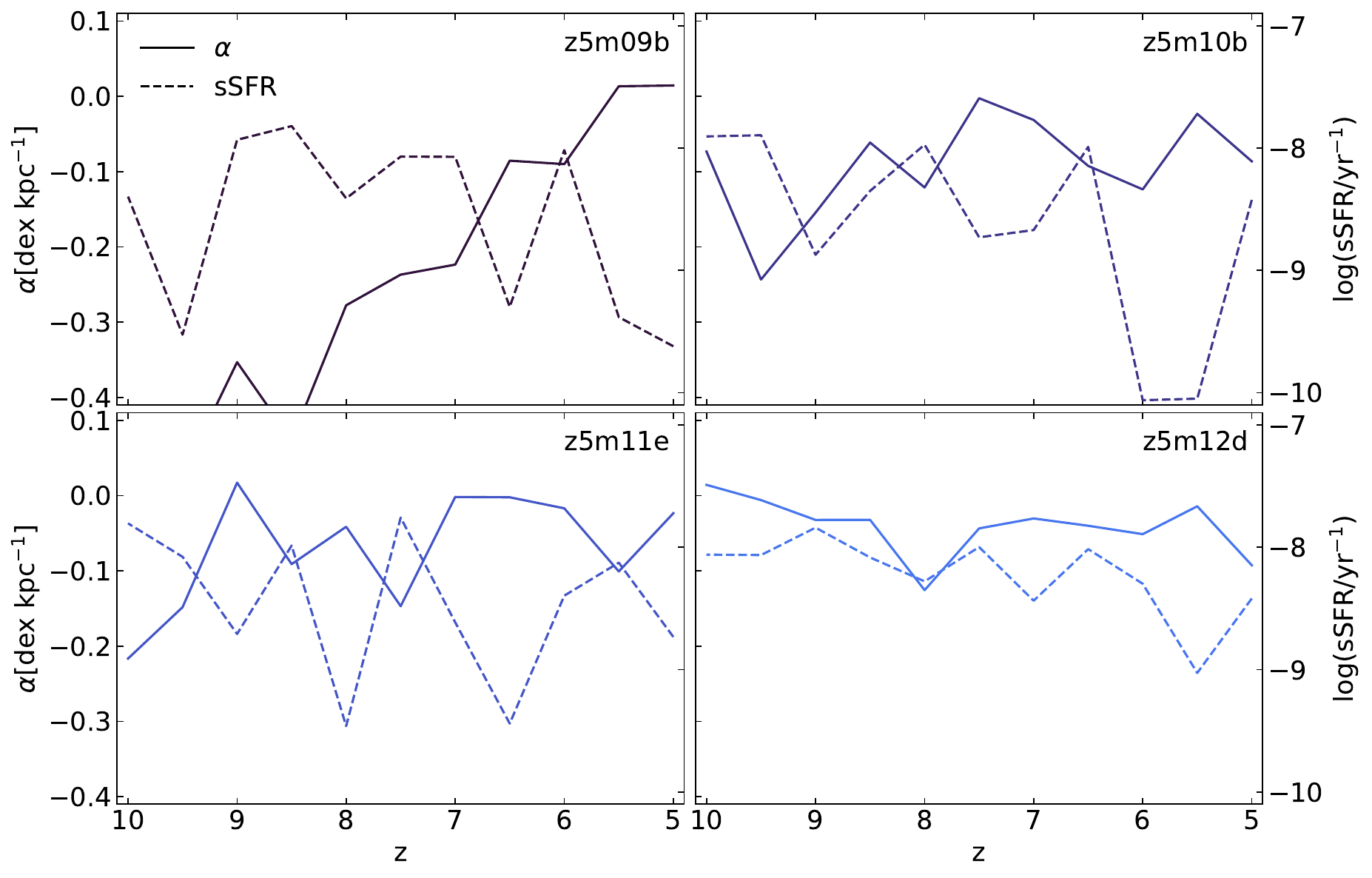}}
 \caption{ 
 \emph{The evolution of four different galaxies.} Low mass galaxies exhibit more pronounced redshift evolution, while the massive ones show a more gradual trend. 
 The overall evolution of these EoR galaxies transitions from steep gradients to flat gradients.
 }
 \label{fig:galaxies}
  \hspace{0.2cm}
\end{figure}

This appendix presents the evolution of four galaxies spanning a range of stellar masses in Fig.~\ref{fig:galaxies}. Higher-mass systems exhibit weaker redshift evolution and overall flatter trends. Given the relatively short cosmic time span covered of these galaxies, the evolution of individual systems can show substantial fluctuations, which may be more apparent in some galaxies than in others, particularly in low-mass galaxies.

\clearpage
\bibliography{sample631}{}

@ARTICLE{Hopkins2018_FIRE2INTRO,
       author = {{Hopkins}, Philip F. and {Wetzel}, Andrew and {Kere{\v{s}}}, Du{\v{s}}an and {Faucher-Gigu{\`e}re}, Claude-Andr{\'e} and {Quataert}, Eliot and {Boylan-Kolchin}, Michael and {Murray}, Norman and {Hayward}, Christopher C. and {Garrison-Kimmel}, Shea and {Hummels}, Cameron and {Feldmann}, Robert and {Torrey}, Paul and {Ma}, Xiangcheng and {Angl{\'e}s-Alc{\'a}zar}, Daniel and {Su}, Kung-Yi and {Orr}, Matthew and {Schmitz}, Denise and {Escala}, Ivanna and {Sanderson}, Robyn and {Grudi{\'c}}, Michael Y. and {Hafen}, Zachary and {Kim}, Ji-Hoon and {Fitts}, Alex and {Bullock}, James S. and {Wheeler}, Coral and {Chan}, T.~K. and {Elbert}, Oliver D. and {Narayanan}, Desika},
        title = "{FIRE-2 simulations: physics versus numerics in galaxy formation}",
      journal = {\mnras},
         year = 2018,
        month = oct,
       volume = {480},
       number = {1},
        pages = {800-863},
          doi = {10.1093/mnras/sty1690},
archivePrefix = {arXiv},
       eprint = {1702.06148},
 primaryClass = {astro-ph.GA},
       adsurl = {https://ui.adsabs.harvard.edu/abs/2018MNRAS.480..800H}
}

@ARTICLE{Hopkins2015_FIRE,
       author = {{Hopkins}, Philip F.},
        title = "{A new class of accurate, mesh-free hydrodynamic simulation methods}",
      journal = {\mnras},
         year = 2015,
        month = jun,
       volume = {450},
       number = {1},
        pages = {53-110},
          doi = {10.1093/mnras/stv195},
archivePrefix = {arXiv},
       eprint = {1409.7395},
 primaryClass = {astro-ph.CO},
       adsurl = {https://ui.adsabs.harvard.edu/abs/2015MNRAS.450...53H}
}

@ARTICLE{Wetzel2023,
       author = {{Wetzel}, Andrew and {Hayward}, Christopher C. and {Sanderson}, Robyn E. and {Ma}, Xiangcheng and {Angl{\'e}s-Alc{\'a}zar}, Daniel and {Feldmann}, Robert and {Chan}, T.~K. and {El-Badry}, Kareem and {Wheeler}, Coral and {Garrison-Kimmel}, Shea and {Nikakhtar}, Farnik and {Panithanpaisal}, Nondh and {Arora}, Arpit and {Gurvich}, Alexander B. and {Samuel}, Jenna and {Sameie}, Omid and {Pandya}, Viraj and {Hafen}, Zachary and {Hummels}, Cameron and {Loebman}, Sarah and {Boylan-Kolchin}, Michael and {Bullock}, James S. and {Faucher-Gigu{\`e}re}, Claude-Andr{\'e} and {Kere{\v{s}}}, Du{\v{s}}an and {Quataert}, Eliot and {Hopkins}, Philip F.},
        title = "{Public Data Release of the FIRE-2 Cosmological Zoom-in Simulations of Galaxy Formation}",
      journal = {\apjs},
         year = 2023,
        month = apr,
       volume = {265},
       number = {2},
          eid = {44},
        pages = {44},
          doi = {10.3847/1538-4365/acb99a},
archivePrefix = {arXiv},
       eprint = {2202.06969},
 primaryClass = {astro-ph.GA},
       adsurl = {https://ui.adsabs.harvard.edu/abs/2023ApJS..265...44W}
}

@ARTICLE{Ma2018_FIRE,
       author = {{Ma}, Xiangcheng and {Hopkins}, Philip F. and {Garrison-Kimmel}, Shea and {Faucher-Gigu{\`e}re}, Claude-Andr{\'e} and {Quataert}, Eliot and {Boylan-Kolchin}, Michael and {Hayward}, Christopher C. and {Feldmann}, Robert and {Kere{\v{s}}}, Du{\v{s}}an},
        title = "{Simulating galaxies in the reionization era with FIRE-2: galaxy scaling relations, stellar mass functions, and luminosity functions}",
      journal = {\mnras},
         year = 2018,
        month = aug,
       volume = {478},
       number = {2},
        pages = {1694-1715},
          doi = {10.1093/mnras/sty1024},
archivePrefix = {arXiv},
       eprint = {1706.06605},
 primaryClass = {astro-ph.GA},
       adsurl = {https://ui.adsabs.harvard.edu/abs/2018MNRAS.478.1694M}
}

@ARTICLE{Ma2019_FIRE,
       author = {{Ma}, Xiangcheng and {Hayward}, Christopher C. and {Casey}, Caitlin M. and {Hopkins}, Philip F. and {Quataert}, Eliot and {Liang}, Lichen and {Faucher-Gigu{\`e}re}, Claude-Andr{\'e} and {Feldmann}, Robert and {Kere{\v{s}}}, Du{\v{s}}an},
        title = "{Dust attenuation, dust emission, and dust temperature in galaxies at z {\ensuremath{\geq}} 5: a view from the FIRE-2 simulations}",
      journal = {\mnras},
         year = 2019,
        month = aug,
       volume = {487},
       number = {2},
        pages = {1844-1864},
          doi = {10.1093/mnras/stz1324},
archivePrefix = {arXiv},
       eprint = {1902.10152},
 primaryClass = {astro-ph.GA},
       adsurl = {https://ui.adsabs.harvard.edu/abs/2019MNRAS.487.1844M}
}

@ARTICLE{Ma2020_FIRE,
       author = {{Ma}, Xiangcheng and {Quataert}, Eliot and {Wetzel}, Andrew and {Hopkins}, Philip F. and {Faucher-Gigu{\`e}re}, Claude-Andr{\'e} and {Kere{\v{s}}}, Du{\v{s}}an},
        title = "{No missing photons for reionization: moderate ionizing photon escape fractions from the FIRE-2 simulations}",
      journal = {\mnras},
         year = 2020,
        month = oct,
       volume = {498},
       number = {2},
        pages = {2001-2017},
          doi = {10.1093/mnras/staa2404},
archivePrefix = {arXiv},
       eprint = {2003.05945},
 primaryClass = {astro-ph.GA},
       adsurl = {https://ui.adsabs.harvard.edu/abs/2020MNRAS.498.2001M}
}

@ARTICLE{Vallini2024_z7,
       author = {{Vallini}, Livia and {Witstok}, Joris and {Sommovigo}, Laura and {Pallottini}, Andrea and {Ferrara}, Andrea and {Carniani}, Stefano and {Kohandel}, Mahsa and {Smit}, Renske and {Gallerani}, Simona and {Gruppioni}, Carlotta},
        title = "{Spatially resolved Kennicutt-Schmidt relation at z {\ensuremath{\approx}} 7 and its connection with the interstellar medium properties}",
      journal = {\mnras},
         year = 2024,
        month = jan,
       volume = {527},
       number = {1},
        pages = {10-22},
          doi = {10.1093/mnras/stad3150},
archivePrefix = {arXiv},
       eprint = {2309.07957},
 primaryClass = {astro-ph.GA},
       adsurl = {https://ui.adsabs.harvard.edu/abs/2024MNRAS.527...10V}
}

@ARTICLE{Venturi2024_z6-8,
       author = {{Venturi}, G. and {Carniani}, S. and {Parlanti}, E. and {Kohandel}, M. and {Curti}, M. and {Pallottini}, A. and {Vallini}, L. and {Arribas}, S. and {Bunker}, A.~J. and {Cameron}, A.~J. and {Castellano}, M. and {Ferrara}, A. and {Fontana}, A. and {Gallerani}, S. and {Gelli}, V. and {Maiolino}, R. and {Ntormousi}, E. and {Pacifici}, C. and {Pentericci}, L. and {Salvadori}, S. and {Vanzella}, E.},
        title = "{Gas-phase metallicity gradients in galaxies at z {\ensuremath{\sim}} 6{\textendash}8}",
      journal = {\aap},
         year = 2024,
        month = nov,
       volume = {691},
          eid = {A19},
        pages = {A19},
          doi = {10.1051/0004-6361/202449855},
archivePrefix = {arXiv},
       eprint = {2403.03977},
 primaryClass = {astro-ph.GA},
       adsurl = {https://ui.adsabs.harvard.edu/abs/2024A&A...691A..19V}
}

@ARTICLE{Tripodi2024_z4-10,
       author = {{Tripodi}, Roberta and {D'Eugenio}, Francesco and {Maiolino}, Roberto and {Curti}, Mirko and {Scholtz}, Jan and {Tacchella}, Sandro and {Marconcini}, Cosimo and {Bunker}, Andrew J. and {Trussler}, James A.~A. and {Cameron}, Alex J. and {Arribas}, Santiago and {Baker}, William M. and {Brada{\v{c}}}, Maru{\v{s}}a and {Carniani}, Stefano and {Charlot}, St{\'e}phane and {Ji}, Xihan and {Ji}, Zhiyuan and {Robertson}, Brant and {{\"U}bler}, Hannah and {Venturi}, Giacomo and {Willmer}, Christopher N.~A. and {Witstok}, Joris},
        title = "{Spatially resolved emission lines in galaxies at 4 {\ensuremath{\leq}} z < 10 from the JADES survey: Evidence for enhanced central star formation}",
      journal = {\aap},
         year = 2024,
        month = dec,
       volume = {692},
          eid = {A184},
        pages = {A184},
          doi = {10.1051/0004-6361/202449980},
archivePrefix = {arXiv},
       eprint = {2403.08431},
 primaryClass = {astro-ph.GA},
       adsurl = {https://ui.adsabs.harvard.edu/abs/2024A&A...692A.184T}
}

@ARTICLE{Arribas2024_z7,
       author = {{Arribas}, Santiago and {Perna}, Michele and {Rodr{\'\i}guez Del Pino}, Bruno and {Lamperti}, Isabella and {D'Eugenio}, Francesco and {P{\'e}rez-Gonz{\'a}lez}, Pablo G. and {Jones}, Gareth C. and {Crespo G{\'o}mez}, Alejandro and {Curti}, Mirko and {Lim}, Seunghwan and {{\'A}lvarez-M{\'a}rquez}, Javier and {Bunker}, Andrew J. and {Carniani}, Stefano and {Charlot}, St{\'e}phane and {Jakobsen}, Peter and {Maiolino}, Roberto and {{\"U}bler}, Hannah and {Willott}, Chris J. and {B{\"o}ker}, Torsten and {Chevallard}, Jacopo and {Circosta}, Chiara and {Cresci}, Giovanni and {Kumari}, Nimisha and {Parlanti}, Eleonora and {Scholtz}, Jan and {Venturi}, Giacomo and {Witstok}, Joris},
        title = "{GA-NIFS: The core of an extremely massive protocluster at the epoch of reionisation probed with JWST/NIRSpec}",
      journal = {\aap},
         year = 2024,
        month = aug,
       volume = {688},
          eid = {A146},
        pages = {A146},
          doi = {10.1051/0004-6361/202348824},
archivePrefix = {arXiv},
       eprint = {2312.00899},
 primaryClass = {astro-ph.GA},
       adsurl = {https://ui.adsabs.harvard.edu/abs/2024A&A...688A.146A}
}

@ARTICLE{Acharyya2024_FOGGIE,
       author = {{Acharyya}, Ayan and {Peeples}, Molly S. and {Tumlinson}, Jason and {O'Shea}, Brian W. and {Lochhaas}, Cassandra and {Wright}, Anna C. and {Simons}, Raymond C. and {Augustin}, Ramona and {Smith}, Britton D. and {Lee}, Eugene Hyeonmin},
        title = "{Figuring Out Gas and Galaxies In Enzo (FOGGIE). VIII. Complex and Stochastic Metallicity Gradients at z > 2}",
      journal = {\apj},
         year = 2025,
        month = feb,
       volume = {979},
       number = {2},
          eid = {129},
        pages = {129},
          doi = {10.3847/1538-4357/ad9dd8},
archivePrefix = {arXiv},
       eprint = {2404.06613},
 primaryClass = {astro-ph.GA},
       adsurl = {https://ui.adsabs.harvard.edu/abs/2025ApJ...979..129A}
}

@ARTICLE{Garcia_EAGLE_TNG,
       author = {{Garcia}, Alex M. and {Torrey}, Paul and {Bhagwat}, Aniket and {Wright}, Ruby J. and {Chen}, Qian-Hui and {Grasha}, Kathryn and {Ridolfo}, Sophia and {Hemler}, Z.~S. and {Sarkar}, Arnab and {Chakraborty}, Priyanka and {Nelson}, Erica J. and {Sanders}, Ryan L. and {Costa}, Tiago and {Vogelsberger}, Mark and {Kewley}, Lisa J. and {Ellison}, Sara L. and {Hernquist}, Lars},
        title = "{Metallicity Gradients in Modern Cosmological Simulations. I. Tension between Smooth Stellar Feedback Models and Observations}",
      journal = {\apj},
         year = 2025,
        month = aug,
       volume = {989},
       number = {2},
          eid = {147},
        pages = {147},
          doi = {10.3847/1538-4357/adea51},
archivePrefix = {arXiv},
       eprint = {2503.03804},
 primaryClass = {astro-ph.GA},
       adsurl = {https://ui.adsabs.harvard.edu/abs/2025ApJ...989..147G}
}

@ARTICLE{Garcia_SPICE,
       author = {{Garcia}, Alex M. and {Torrey}, Paul and {Bhagwat}, Aniket and {Shen}, Xuejian and {Vogelsberger}, Mark and {McClymont}, William and {Nagarajan-Swenson}, Jaya and {Ridolfo}, Sophia G. and {Zhu}, Peixin and {Zimmerman}, Dhruv T. and {Zier}, Oliver and {Biddle}, Sarah and {Sarkar}, Arnab and {Chakraborty}, Priyanka and {Wright}, Ruby J. and {Grasha}, Kathryn and {Costa}, Tiago and {Keating}, Laura and {Kannan}, Rahul and {Smith}, Aaron and {Garaldi}, Enrico and {Puchwein}, Ewald and {Ciardi}, Benedetta and {Hernquist}, Lars and {Kewley}, Lisa J.},
        title = "{Metallicity Gradients in Modern Cosmological Simulations II: The Role of Bursty Versus Smooth Feedback at High-Redshift}",
      journal = {arXiv e-prints},
         year = 2025,
        month = oct,
          eid = {arXiv:2510.26877},
        pages = {arXiv:2510.26877},
          doi = {10.48550/arXiv.2510.26877},
archivePrefix = {arXiv},
       eprint = {2510.26877},
 primaryClass = {astro-ph.GA},
       adsurl = {https://ui.adsabs.harvard.edu/abs/2025arXiv251026877G}
}

@ARTICLE{Ma2018_FIRE_size,
       author = {{Ma}, Xiangcheng and {Hopkins}, Philip F. and {Boylan-Kolchin}, Michael and {Faucher-Gigu{\`e}re}, Claude-Andr{\'e} and {Quataert}, Eliot and {Feldmann}, Robert and {Garrison-Kimmel}, Shea and {Hayward}, Christopher C. and {Kere{\v{s}}}, Du{\v{s}}an and {Wetzel}, Andrew},
        title = "{Simulating galaxies in the reionization era with FIRE-2: morphologies and sizes}",
      journal = {\mnras},
         year = 2018,
        month = jun,
       volume = {477},
       number = {1},
        pages = {219-229},
          doi = {10.1093/mnras/sty684},
archivePrefix = {arXiv},
       eprint = {1710.00008},
 primaryClass = {astro-ph.GA},
       adsurl = {https://ui.adsabs.harvard.edu/abs/2018MNRAS.477..219M}
}

@ARTICLE{Ma2017_FIRE_gra,
       author = {{Ma}, Xiangcheng and {Hopkins}, Philip F. and {Feldmann}, Robert and {Torrey}, Paul and {Faucher-Gigu{\`e}re}, Claude-Andr{\'e} and {Kere{\v{s}}}, Du{\v{s}}an},
        title = "{Why do high-redshift galaxies show diverse gas-phase metallicity gradients?}",
      journal = {\mnras},
         year = 2017,
        month = apr,
       volume = {466},
       number = {4},
        pages = {4780-4794},
          doi = {10.1093/mnras/stx034},
archivePrefix = {arXiv},
       eprint = {1610.03498},
 primaryClass = {astro-ph.GA},
       adsurl = {https://ui.adsabs.harvard.edu/abs/2017MNRAS.466.4780M}
}

@ARTICLE{Sun2024_z04_3,
       author = {{Sun}, Xunda and {Wang}, Xin and {Ma}, Xiangcheng and {Wang}, Kai and {Wetzel}, Andrew and {Faucher-Gigu{\`e}re}, Claude-Andr{\'e} and {Hopkins}, Philip F. and {Kere{\v{s}}}, Du{\v{s}}an and {Graf}, Russell L. and {Marszewski}, Andrew and {Stern}, Jonathan and {Sun}, Guochao and {Sun}, Lei and {Thyme}, Keyer},
        title = "{The Physical Origin of Positive Metallicity Radial Gradients in High-redshift Galaxies: Insights from the FIRE-2 Cosmological Hydrodynamic Simulations}",
      journal = {\apj},
         year = 2025,
        month = jun,
       volume = {986},
       number = {2},
          eid = {179},
        pages = {179},
          doi = {10.3847/1538-4357/addab5},
archivePrefix = {arXiv},
       eprint = {2409.09290},
 primaryClass = {astro-ph.GA},
       adsurl = {https://ui.adsabs.harvard.edu/abs/2025ApJ...986..179S}
}

@ARTICLE{Orr2023,
       author = {{Orr}, Matthew E. and {Burkhart}, Blakesley and {Wetzel}, Andrew and {Hopkins}, Philip F. and {Escala}, Ivanna A. and {Strom}, Allison L. and {Goldsmith}, Paul F. and {Pineda}, Jorge L. and {Hayward}, Christopher C. and {Loebman}, Sarah R.},
        title = "{Spiral arms are metal freeways: azimuthal gas-phase metallicity variations in flocculent discs in the FIRE-2 cosmological zoom-in simulations}",
      journal = {\mnras},
         year = 2023,
        month = may,
       volume = {521},
       number = {3},
        pages = {3708-3726},
          doi = {10.1093/mnras/stad676},
archivePrefix = {arXiv},
       eprint = {2209.14159},
 primaryClass = {astro-ph.GA},
       adsurl = {https://ui.adsabs.harvard.edu/abs/2023MNRAS.521.3708O}
}

@ARTICLE{Graf2024,
       author = {{Graf}, Russell L. and {Wetzel}, Andrew and {Bailin}, Jeremy and {Orr}, Matthew E.},
        title = "{Inside-out versus Upside-down: The Origin and Evolution of Metallicity Radial Gradients in FIRE Simulations of Milky Way-mass Galaxies and the Essential Role of Gas Mixing}",
      journal = {arXiv e-prints},
         year = 2024,
        month = oct,
          eid = {arXiv:2410.21377},
        pages = {arXiv:2410.21377},
          doi = {10.48550/arXiv.2410.21377},
archivePrefix = {arXiv},
       eprint = {2410.21377},
 primaryClass = {astro-ph.GA},
       adsurl = {https://ui.adsabs.harvard.edu/abs/2024arXiv241021377G}
}

@ARTICLE{Graf2025,
       author = {{Graf}, Russell L. and {Wetzel}, Andrew and {Bellardini}, Matthew A. and {Bailin}, Jeremy},
        title = "{Spatial Variations of Stellar Elemental Abundances in FIRE Simulations of Milky Way-mass Galaxies: Patterns Today Mostly Reflect Those at Formation}",
      journal = {\apj},
         year = 2025,
        month = mar,
       volume = {981},
       number = {1},
          eid = {47},
        pages = {47},
          doi = {10.3847/1538-4357/adacd7},
archivePrefix = {arXiv},
       eprint = {2402.15614},
 primaryClass = {astro-ph.GA},
       adsurl = {https://ui.adsabs.harvard.edu/abs/2025ApJ...981...47G}
}

@ARTICLE{Bellardini2021,
       author = {{Bellardini}, Matthew A. and {Wetzel}, Andrew and {Loebman}, Sarah R. and {Faucher-Gigu{\`e}re}, Claude-Andr{\'e} and {Ma}, Xiangcheng and {Feldmann}, Robert},
        title = "{3D gas-phase elemental abundances across the formation histories of Milky Way-mass galaxies in the FIRE simulations: initial conditions for chemical tagging}",
      journal = {\mnras},
         year = 2021,
        month = aug,
       volume = {505},
       number = {3},
        pages = {4586-4607},
          doi = {10.1093/mnras/stab1606},
archivePrefix = {arXiv},
       eprint = {2102.06220},
 primaryClass = {astro-ph.GA},
       adsurl = {https://ui.adsabs.harvard.edu/abs/2021MNRAS.505.4586B}
}

@ARTICLE{Bellardini2022,
       author = {{Bellardini}, Matthew A. and {Wetzel}, Andrew and {Loebman}, Sarah R. and {Bailin}, Jeremy},
        title = "{3D elemental abundances of stars at formation across the histories of Milky Way-mass galaxies in the FIRE simulations}",
      journal = {\mnras},
         year = 2022,
        month = aug,
       volume = {514},
       number = {3},
        pages = {4270-4289},
          doi = {10.1093/mnras/stac1637},
archivePrefix = {arXiv},
       eprint = {2203.03653},
 primaryClass = {astro-ph.GA},
       adsurl = {https://ui.adsabs.harvard.edu/abs/2022MNRAS.514.4270B}
}

@ARTICLE{Wiersma2009_elem,
       author = {{Wiersma}, Robert P.~C. and {Schaye}, Joop and {Smith}, Britton D.},
        title = "{The effect of photoionization on the cooling rates of enriched, astrophysical plasmas}",
      journal = {\mnras},
         year = 2009,
        month = feb,
       volume = {393},
       number = {1},
        pages = {99-107},
          doi = {10.1111/j.1365-2966.2008.14191.x},
archivePrefix = {arXiv},
       eprint = {0807.3748},
 primaryClass = {astro-ph},
       adsurl = {https://ui.adsabs.harvard.edu/abs/2009MNRAS.393...99W}
}

@ARTICLE{Leitherer1999,
       author = {{Leitherer}, Claus and {Schaerer}, Daniel and {Goldader}, Jeffrey D. and {Delgado}, Rosa M. Gonz{\'a}lez and {Robert}, Carmelle and {Kune}, Denis Foo and {de Mello}, Du{\'\i}lia F. and {Devost}, Daniel and {Heckman}, Timothy M.},
        title = "{Starburst99: Synthesis Models for Galaxies with Active Star Formation}",
      journal = {\apjs},
         year = 1999,
        month = jul,
       volume = {123},
       number = {1},
        pages = {3-40},
          doi = {10.1086/313233},
archivePrefix = {arXiv},
       eprint = {astro-ph/9902334},
 primaryClass = {astro-ph},
       adsurl = {https://ui.adsabs.harvard.edu/abs/1999ApJS..123....3L}
}

@ARTICLE{Kroupa2001,
       author = {{Kroupa}, Pavel},
        title = "{On the variation of the initial mass function}",
      journal = {\mnras},
         year = 2001,
        month = apr,
       volume = {322},
       number = {2},
        pages = {231-246},
          doi = {10.1046/j.1365-8711.2001.04022.x},
archivePrefix = {arXiv},
       eprint = {astro-ph/0009005},
 primaryClass = {astro-ph},
       adsurl = {https://ui.adsabs.harvard.edu/abs/2001MNRAS.322..231K}
}

@ARTICLE{Bouwens2004,
       author = {{Bouwens}, R.~J. and {Illingworth}, G.~D. and {Blakeslee}, J.~P. and {Broadhurst}, T.~J. and {Franx}, M.},
        title = "{Galaxy Size Evolution at High Redshift and Surface Brightness Selection Effects: Constraints from the Hubble Ultra Deep Field}",
      journal = {\apjl},
         year = 2004,
        month = aug,
       volume = {611},
       number = {1},
        pages = {L1-L4},
          doi = {10.1086/423786},
archivePrefix = {arXiv},
       eprint = {astro-ph/0406562},
 primaryClass = {astro-ph},
       adsurl = {https://ui.adsabs.harvard.edu/abs/2004ApJ...611L...1B}
}

@ARTICLE{Leethochawalit2016,
       author = {{Leethochawalit}, Nicha and {Jones}, Tucker A. and {Ellis}, Richard S. and {Stark}, Daniel P. and {Richard}, Johan and {Zitrin}, Adi and {Auger}, Matthew},
        title = "{A Keck Adaptive Optics Survey of a Representative Sample of Gravitationally Lensed Star-forming Galaxies: High Spatial Resolution Studies of Kinematics and Metallicity Gradients}",
      journal = {\apj},
         year = 2016,
        month = apr,
       volume = {820},
       number = {2},
          eid = {84},
        pages = {84},
          doi = {10.3847/0004-637X/820/2/84},
archivePrefix = {arXiv},
       eprint = {1509.01279},
 primaryClass = {astro-ph.GA},
       adsurl = {https://ui.adsabs.harvard.edu/abs/2016ApJ...820...84L}
}

@ARTICLE{Searle1971,
       author = {{Searle}, Leonard},
        title = "{Evidence for Composition Gradients across the Disks of Spiral Galaxies}",
      journal = {\apj},
         year = 1971,
        month = sep,
       volume = {168},
        pages = {327},
          doi = {10.1086/151090},
       adsurl = {https://ui.adsabs.harvard.edu/abs/1971ApJ...168..327S}
}

@ARTICLE{Gallazzi2005,
       author = {{Gallazzi}, Anna and {Charlot}, St{\'e}phane and {Brinchmann}, Jarle and {White}, Simon D.~M. and {Tremonti}, Christy A.},
        title = "{The ages and metallicities of galaxies in the local universe}",
      journal = {\mnras},
         year = 2005,
        month = sep,
       volume = {362},
       number = {1},
        pages = {41-58},
          doi = {10.1111/j.1365-2966.2005.09321.x},
archivePrefix = {arXiv},
       eprint = {astro-ph/0506539},
 primaryClass = {astro-ph},
       adsurl = {https://ui.adsabs.harvard.edu/abs/2005MNRAS.362...41G}
}

@ARTICLE{Kirby2013,
       author = {{Kirby}, Evan N. and {Cohen}, Judith G. and {Guhathakurta}, Puragra and {Cheng}, Lucy and {Bullock}, James S. and {Gallazzi}, Anna},
        title = "{The Universal Stellar Mass-Stellar Metallicity Relation for Dwarf Galaxies}",
      journal = {\apj},
         year = 2013,
        month = dec,
       volume = {779},
       number = {2},
          eid = {102},
        pages = {102},
          doi = {10.1088/0004-637X/779/2/102},
archivePrefix = {arXiv},
       eprint = {1310.0814},
 primaryClass = {astro-ph.GA},
       adsurl = {https://ui.adsabs.harvard.edu/abs/2013ApJ...779..102K}
}

@ARTICLE{Tremonti2004,
       author = {{Tremonti}, Christy A. and {Heckman}, Timothy M. and {Kauffmann}, Guinevere and {Brinchmann}, Jarle and {Charlot}, St{\'e}phane and {White}, Simon D.~M. and {Seibert}, Mark and {Peng}, Eric W. and {Schlegel}, David J. and {Uomoto}, Alan and {Fukugita}, Masataka and {Brinkmann}, Jon},
        title = "{The Origin of the Mass-Metallicity Relation: Insights from 53,000 Star-forming Galaxies in the Sloan Digital Sky Survey}",
      journal = {\apj},
         year = 2004,
        month = oct,
       volume = {613},
       number = {2},
        pages = {898-913},
          doi = {10.1086/423264},
archivePrefix = {arXiv},
       eprint = {astro-ph/0405537},
 primaryClass = {astro-ph},
       adsurl = {https://ui.adsabs.harvard.edu/abs/2004ApJ...613..898T}
}

@ARTICLE{Lee2006,
       author = {{Lee}, Henry and {Skillman}, Evan D. and {Cannon}, John M. and {Jackson}, Dale C. and {Gehrz}, Robert D. and {Polomski}, Elisha F. and {Woodward}, Charles E.},
        title = "{On Extending the Mass-Metallicity Relation of Galaxies by 2.5 Decades in Stellar Mass}",
      journal = {\apj},
         year = 2006,
        month = aug,
       volume = {647},
       number = {2},
        pages = {970-983},
          doi = {10.1086/505573},
archivePrefix = {arXiv},
       eprint = {astro-ph/0605036},
 primaryClass = {astro-ph},
       adsurl = {https://ui.adsabs.harvard.edu/abs/2006ApJ...647..970L}
}

@ARTICLE{King2015,
       author = {{King}, Andrew and {Pounds}, Ken},
        title = "{Powerful Outflows and Feedback from Active Galactic Nuclei}",
      journal = {\araa},
         year = 2015,
        month = aug,
       volume = {53},
        pages = {115-154},
          doi = {10.1146/annurev-astro-082214-122316},
archivePrefix = {arXiv},
       eprint = {1503.05206},
 primaryClass = {astro-ph.GA},
       adsurl = {https://ui.adsabs.harvard.edu/abs/2015ARA&A..53..115K}
}

@ARTICLE{Thielemann2017,
       author = {{Thielemann}, F. -K. and {Eichler}, M. and {Panov}, I.~V. and {Wehmeyer}, B.},
        title = "{Neutron Star Mergers and Nucleosynthesis of Heavy Elements}",
      journal = {Annual Review of Nuclear and Particle Science},
         year = 2017,
        month = oct,
       volume = {67},
        pages = {253-274},
          doi = {10.1146/annurev-nucl-101916-123246},
archivePrefix = {arXiv},
       eprint = {1710.02142},
 primaryClass = {astro-ph.HE},
       adsurl = {https://ui.adsabs.harvard.edu/abs/2017ARNPS..67..253T}
}

@ARTICLE{Maiolino2019,
       author = {{Maiolino}, R. and {Mannucci}, F.},
        title = "{De re metallica: the cosmic chemical evolution of galaxies}",
      journal = {\aapr},
         year = 2019,
        month = feb,
       volume = {27},
       number = {1},
          eid = {3},
        pages = {3},
          doi = {10.1007/s00159-018-0112-2},
archivePrefix = {arXiv},
       eprint = {1811.09642},
 primaryClass = {astro-ph.GA},
       adsurl = {https://ui.adsabs.harvard.edu/abs/2019A&ARv..27....3M}
}

@ARTICLE{Wangk2023,
       author = {{Wang}, Kai and {Wang}, Xin and {Chen}, Yangyao},
        title = "{Environmental Dependence of the Mass-Metallicity Relation in Cosmological Hydrodynamical Simulations}",
      journal = {\apj},
         year = 2023,
        month = jul,
       volume = {951},
       number = {1},
          eid = {66},
        pages = {66},
          doi = {10.3847/1538-4357/acd633},
archivePrefix = {arXiv},
       eprint = {2305.08161},
 primaryClass = {astro-ph.GA},
       adsurl = {https://ui.adsabs.harvard.edu/abs/2023ApJ...951...66W}
}

@ARTICLE{Zaritsky1994,
       author = {{Zaritsky}, Dennis and {Kennicutt}, Robert C., Jr. and {Huchra}, John P.},
        title = "{H II Regions and the Abundance Properties of Spiral Galaxies}",
      journal = {\apj},
         year = 1994,
        month = jan,
       volume = {420},
        pages = {87},
          doi = {10.1086/173544},
       adsurl = {https://ui.adsabs.harvard.edu/abs/1994ApJ...420...87Z}
}

@ARTICLE{vanZee1998,
       author = {{van Zee}, Liese and {Salzer}, John J. and {Haynes}, Martha P. and {O'Donoghue}, Aileen A. and {Balonek}, Thomas J.},
        title = "{Spectroscopy of Outlying H II Regions in Spiral Galaxies: Abundances and Radial Gradients}",
      journal = {\aj},
         year = 1998,
        month = dec,
       volume = {116},
       number = {6},
        pages = {2805-2833},
          doi = {10.1086/300647},
archivePrefix = {arXiv},
       eprint = {astro-ph/9808315},
 primaryClass = {astro-ph},
       adsurl = {https://ui.adsabs.harvard.edu/abs/1998AJ....116.2805V}
}

@ARTICLE{Stott2014,
       author = {{Stott}, John P. and {Sobral}, David and {Swinbank}, A.~M. and {Smail}, Ian and {Bower}, Richard and {Best}, Philip N. and {Sharples}, Ray M. and {Geach}, James E. and {Matthee}, Jorryt},
        title = "{A relationship between specific star formation rate and metallicity gradient within z {\ensuremath{\sim}} 1 galaxies from KMOS-HiZELS}",
      journal = {\mnras},
         year = 2014,
        month = sep,
       volume = {443},
       number = {3},
        pages = {2695-2704},
          doi = {10.1093/mnras/stu1343},
archivePrefix = {arXiv},
       eprint = {1407.1047},
 primaryClass = {astro-ph.GA},
       adsurl = {https://ui.adsabs.harvard.edu/abs/2014MNRAS.443.2695S}
}

@ARTICLE{Carton2018,
       author = {{Carton}, David and {Brinchmann}, Jarle and {Contini}, Thierry and {Epinat}, Beno{\^\i}t and {Finley}, Hayley and {Richard}, Johan and {Patr{\'\i}cio}, Vera and {Schaye}, Joop and {Nanayakkara}, Themiya and {Weilbacher}, Peter M. and {Wisotzki}, Lutz},
        title = "{First gas-phase metallicity gradients of 0.1 {\ensuremath{\lesssim}} z {\ensuremath{\lesssim}} 0.8 galaxies with MUSE}",
      journal = {\mnras},
         year = 2018,
        month = aug,
       volume = {478},
       number = {4},
        pages = {4293-4316},
          doi = {10.1093/mnras/sty1343},
archivePrefix = {arXiv},
       eprint = {1805.08131},
 primaryClass = {astro-ph.GA},
       adsurl = {https://ui.adsabs.harvard.edu/abs/2018MNRAS.478.4293C}
}

@ARTICLE{Wuyts2016,
       author = {{Wuyts}, Eva and {Wisnioski}, Emily and {Fossati}, Matteo and {F{\"o}rster Schreiber}, Natascha M. and {Genzel}, Reinhard and {Davies}, Ric and {Mendel}, J. Trevor and {Naab}, Thorsten and {R{\"o}ttgers}, Bernhard and {Wilman}, David J. and {Wuyts}, Stijn and {Bandara}, Kaushala and {Beifiori}, Alessandra and {Belli}, Sirio and {Bender}, Ralf and {Brammer}, Gabriel B. and {Burkert}, Andreas and {Chan}, Jeffrey and {Galametz}, Audrey and {Kulkarni}, Sandesh K. and {Lang}, Philipp and {Lutz}, Dieter and {Momcheva}, Ivelina G. and {Nelson}, Erica J. and {Rosario}, David and {Saglia}, Roberto P. and {Seitz}, Stella and {Tacconi}, Linda J. and {Tadaki}, Ken-ichi and {{\"U}bler}, Hannah and {van Dokkum}, Pieter},
        title = "{The Evolution of Metallicity and Metallicity Gradients from z = 2.7 to 0.6 with KMOS$^{3D}$}",
      journal = {\apj},
         year = 2016,
        month = aug,
       volume = {827},
       number = {1},
          eid = {74},
        pages = {74},
          doi = {10.3847/0004-637X/827/1/74},
archivePrefix = {arXiv},
       eprint = {1603.01139},
 primaryClass = {astro-ph.GA},
       adsurl = {https://ui.adsabs.harvard.edu/abs/2016ApJ...827...74W}
}

@ARTICLE{Cresci2010,
       author = {{Cresci}, G. and {Mannucci}, F. and {Maiolino}, R. and {Marconi}, A. and {Gnerucci}, A. and {Magrini}, L.},
        title = "{Gas accretion as the origin of chemical abundance gradients in distant galaxies}",
      journal = {\nat},
         year = 2010,
        month = oct,
       volume = {467},
       number = {7317},
        pages = {811-813},
          doi = {10.1038/nature09451},
archivePrefix = {arXiv},
       eprint = {1010.2534},
 primaryClass = {astro-ph.CO},
       adsurl = {https://ui.adsabs.harvard.edu/abs/2010Natur.467..811C}
}

@ARTICLE{Wang2019,
       author = {{Wang}, Xin and {Jones}, Tucker A. and {Treu}, Tommaso and {Hirtenstein}, Jessie and {Brammer}, Gabriel B. and {Daddi}, Emanuele and {Meng}, Xiao-Lei and {Morishita}, Takahiro and {Abramson}, Louis E. and {Henry}, Alaina L. and {Peng}, Ying-jie and {Schmidt}, Kasper B. and {Sharon}, Keren and {Trenti}, Michele and {Vulcani}, Benedetta},
        title = "{Discovery of Strongly Inverted Metallicity Gradients in Dwarf Galaxies at z {\ensuremath{\sim}} 2}",
      journal = {\apj},
         year = 2019,
        month = sep,
       volume = {882},
       number = {2},
          eid = {94},
        pages = {94},
          doi = {10.3847/1538-4357/ab3861},
archivePrefix = {arXiv},
       eprint = {1808.08800},
 primaryClass = {astro-ph.GA},
       adsurl = {https://ui.adsabs.harvard.edu/abs/2019ApJ...882...94W}
}

@ARTICLE{Ju2025,
       author = {{Ju}, Mengting and {Wang}, Xin and {Jones}, Tucker and {Bari{\v{s}}i{\'c}}, Ivana and {Nanayakkara}, Themiya and {Bundy}, Kevin and {Faucher-Gigu{\`e}re}, Claude-Andr{\'e} and {Feng}, Shuai and {Glazebrook}, Karl and {Henry}, Alaina and {Malkan}, Matthew A. and {Obreschkow}, Danail and {Roy}, Namrata and {Sanders}, Ryan L. and {Sun}, Xunda and {Treu}, Tommaso and {Zhou}, Qianqiao},
        title = "{MSA-3D: Metallicity Gradients in Galaxies at z {\ensuremath{\sim}} 1 with JWST/NIRSpec Slit-stepping Spectroscopy}",
      journal = {\apjl},
         year = 2025,
        month = jan,
       volume = {978},
       number = {2},
          eid = {L39},
        pages = {L39},
          doi = {10.3847/2041-8213/ada150},
archivePrefix = {arXiv},
       eprint = {2409.01616},
 primaryClass = {astro-ph.GA},
       adsurl = {https://ui.adsabs.harvard.edu/abs/2025ApJ...978L..39J}
}

@ARTICLE{Nomoto2013,
       author = {{Nomoto}, Ken'ichi and {Kobayashi}, Chiaki and {Tominaga}, Nozomu},
        title = "{Nucleosynthesis in Stars and the Chemical Enrichment of Galaxies}",
      journal = {\araa},
         year = 2013,
        month = aug,
       volume = {51},
       number = {1},
        pages = {457-509},
          doi = {10.1146/annurev-astro-082812-140956},
       adsurl = {https://ui.adsabs.harvard.edu/abs/2013ARA&A..51..457N}
}

@ARTICLE{Jones2010,
       author = {{Jones}, T.~A. and {Swinbank}, A.~M. and {Ellis}, R.~S. and {Richard}, J. and {Stark}, D.~P.},
        title = "{Resolved spectroscopy of gravitationally lensed galaxies: recovering coherent velocity fields in subluminous z \raisebox{-0.5ex}\textasciitilde 2-3 galaxies}",
      journal = {\mnras},
         year = 2010,
        month = may,
       volume = {404},
       number = {3},
        pages = {1247-1262},
          doi = {10.1111/j.1365-2966.2010.16378.x},
archivePrefix = {arXiv},
       eprint = {0910.4488},
 primaryClass = {astro-ph.CO},
       adsurl = {https://ui.adsabs.harvard.edu/abs/2010MNRAS.404.1247J}
}

@ARTICLE{Swinbank2012,
       author = {{Swinbank}, A.~M. and {Sobral}, D. and {Smail}, Ian and {Geach}, J.~E. and {Best}, P.~N. and {McCarthy}, I.~G. and {Crain}, R.~A. and {Theuns}, T.},
        title = "{The properties of the star-forming interstellar medium at z = 0.84-2.23 from HiZELS: mapping the internal dynamics and metallicity gradients in high-redshift disc galaxies}",
      journal = {\mnras},
         year = 2012,
        month = oct,
       volume = {426},
       number = {2},
        pages = {935-950},
          doi = {10.1111/j.1365-2966.2012.21774.x},
archivePrefix = {arXiv},
       eprint = {1209.1395},
 primaryClass = {astro-ph.CO},
       adsurl = {https://ui.adsabs.harvard.edu/abs/2012MNRAS.426..935S}
}

@ARTICLE{Yuan2011,
       author = {{Yuan}, T. -T. and {Kewley}, L.~J. and {Swinbank}, A.~M. and {Richard}, J. and {Livermore}, R.~C.},
        title = "{Metallicity Gradient of a Lensed Face-on Spiral Galaxy at Redshift 1.49}",
      journal = {\apjl},
         year = 2011,
        month = may,
       volume = {732},
       number = {1},
          eid = {L14},
        pages = {L14},
          doi = {10.1088/2041-8205/732/1/L14},
archivePrefix = {arXiv},
       eprint = {1103.3277},
 primaryClass = {astro-ph.CO},
       adsurl = {https://ui.adsabs.harvard.edu/abs/2011ApJ...732L..14Y}
}

@ARTICLE{Knollmann2009,
       author = {{Knollmann}, Steffen R. and {Knebe}, Alexander},
        title = "{AHF: Amiga's Halo Finder}",
      journal = {\apjs},
         year = 2009,
        month = jun,
       volume = {182},
       number = {2},
        pages = {608-624},
          doi = {10.1088/0067-0049/182/2/608},
archivePrefix = {arXiv},
       eprint = {0904.3662},
 primaryClass = {astro-ph.CO},
       adsurl = {https://ui.adsabs.harvard.edu/abs/2009ApJS..182..608K}
}

@ARTICLE{Girard2020,
       author = {{Girard}, M. and {Mason}, C.~A. and {Fontana}, A. and {Dessauges-Zavadsky}, M. and {Morishita}, T. and {Amor{\'\i}n}, R. and {Fisher}, D.~B. and {Jones}, T. and {Schaerer}, D. and {Schmidt}, K.~B. and {Treu}, T. and {Vulcani}, B.},
        title = "{The KMOS Lens-Amplified Spectroscopic Survey (KLASS): kinematics and clumpiness of low-mass galaxies at cosmic noon}",
      journal = {\mnras},
         year = 2020,
        month = sep,
       volume = {497},
       number = {1},
        pages = {173-191},
          doi = {10.1093/mnras/staa1907},
archivePrefix = {arXiv},
       eprint = {2006.14633},
 primaryClass = {astro-ph.GA},
       adsurl = {https://ui.adsabs.harvard.edu/abs/2020MNRAS.497..173G}
}

@ARTICLE{Kassin2012,
       author = {{Kassin}, Susan A. and {Weiner}, Benjamin J. and {Faber}, S.~M. and {Gardner}, Jonathan P. and {Willmer}, C.~N.~A. and {Coil}, Alison L. and {Cooper}, Michael C. and {Devriendt}, Julien and {Dutton}, Aaron A. and {Guhathakurta}, Puragra and {Koo}, David C. and {Metevier}, A.~J. and {Noeske}, Kai G. and {Primack}, Joel R.},
        title = "{The Epoch of Disk Settling: z \raisebox{-0.5ex}\textasciitilde 1 to Now}",
      journal = {\apj},
         year = 2012,
        month = oct,
       volume = {758},
       number = {2},
          eid = {106},
        pages = {106},
          doi = {10.1088/0004-637X/758/2/106},
archivePrefix = {arXiv},
       eprint = {1207.7072},
 primaryClass = {astro-ph.CO},
       adsurl = {https://ui.adsabs.harvard.edu/abs/2012ApJ...758..106K}
}

@ARTICLE{Fitts2017,
       author = {{Fitts}, Alex and {Boylan-Kolchin}, Michael and {Elbert}, Oliver D. and {Bullock}, James S. and {Hopkins}, Philip F. and {O{\~n}orbe}, Jose and {Wetzel}, Andrew and {Wheeler}, Coral and {Faucher-Gigu{\`e}re}, Claude-Andr{\'e} and {Kere{\v{s}}}, Du{\v{s}}an and {Skillman}, Evan D. and {Weisz}, Daniel R.},
        title = "{fire in the field: simulating the threshold of galaxy formation}",
      journal = {\mnras},
         year = 2017,
        month = nov,
       volume = {471},
       number = {3},
        pages = {3547-3562},
          doi = {10.1093/mnras/stx1757},
archivePrefix = {arXiv},
       eprint = {1611.02281},
 primaryClass = {astro-ph.GA},
       adsurl = {https://ui.adsabs.harvard.edu/abs/2017MNRAS.471.3547F}
}

@ARTICLE{Orr2017,
       author = {{Orr}, Matthew E. and {Hayward}, Christopher C. and {Nelson}, Erica J. and {Hopkins}, Philip F. and {Faucher-Gigu{\`e}re}, Claude-Andr{\'e} and {Kere{\v{s}}}, Du{\v{s}}an and {Chan}, T.~K. and {Schmitz}, Denise M. and {Miller}, Tim B.},
        title = "{Stacked Star Formation Rate Profiles of Bursty Galaxies Exhibit {\textquotedblleft}Coherent{\textquotedblright} Star Formation}",
      journal = {\apjl},
         year = 2017,
        month = nov,
       volume = {849},
       number = {1},
          eid = {L2},
        pages = {L2},
          doi = {10.3847/2041-8213/aa8f93},
archivePrefix = {arXiv},
       eprint = {1709.10099},
 primaryClass = {astro-ph.GA},
       adsurl = {https://ui.adsabs.harvard.edu/abs/2017ApJ...849L...2O}
}

@ARTICLE{Orr2018,
       author = {{Orr}, Matthew E. and {Hayward}, Christopher C. and {Hopkins}, Philip F. and {Chan}, T.~K. and {Faucher-Gigu{\`e}re}, Claude-Andr{\'e} and {Feldmann}, Robert and {Kere{\v{s}}}, Du{\v{s}}an and {Murray}, Norman and {Quataert}, Eliot},
        title = "{What FIREs up star formation: the emergence of the Kennicutt-Schmidt law from feedback}",
      journal = {\mnras},
         year = 2018,
        month = aug,
       volume = {478},
       number = {3},
        pages = {3653-3673},
          doi = {10.1093/mnras/sty1241},
archivePrefix = {arXiv},
       eprint = {1701.01788},
 primaryClass = {astro-ph.GA},
       adsurl = {https://ui.adsabs.harvard.edu/abs/2018MNRAS.478.3653O}
}

@ARTICLE{Li2025,
       author = {{Li}, Zihao and {Cai}, Zheng and {Wang}, Xin and {Li}, Zhaozhou and {Dekel}, Avishai and {Sarkar}, Kartick C. and {Ba{\~n}ados}, Eduardo and {Bian}, Fuyan and {Bhowmick}, Aklant K. and {Blecha}, Laura and {Bosman}, Sarah E.~I. and {Champagne}, Jaclyn B. and {Fan}, Xiaohui and {Golden-Marx}, Emmet and {Jun}, Hyunsung D. and {Li}, Mingyu and {Lin}, Xiaojing and {Liu}, Weizhe and {Sun}, Fengwu and {Trebitsch}, Maxime and {Walter}, Fabian and {Wang}, Feige and {Wu}, Yunjing and {Yang}, Jinyi and {Zhang}, Huanian and {Zhang}, Shiwu and {Zhuang}, Mingyang and {Zou}, Siwei},
        title = "{A 13 Billion Year View of Galaxy Growth: Metallicity Gradient Evolution from the Local Universe to z = 9 with JWST and Archival Surveys}",
      journal = {\apjs},
         year = 2025,
        month = oct,
       volume = {280},
       number = {2},
          eid = {62},
        pages = {62},
          doi = {10.3847/1538-4365/adfa70},
archivePrefix = {arXiv},
       eprint = {2506.12129},
 primaryClass = {astro-ph.GA},
       adsurl = {https://ui.adsabs.harvard.edu/abs/2025ApJS..280...62L}
}

@ARTICLE{Wang2017,
       author = {{Wang}, Xin and {Jones}, Tucker A. and {Treu}, Tommaso and {Morishita}, Takahiro and {Abramson}, Louis E. and {Brammer}, Gabriel B. and {Huang}, Kuang-Han and {Malkan}, Matthew A. and {Schmidt}, Kasper B. and {Fontana}, Adriano and {Grillo}, Claudio and {Henry}, Alaina L. and {Karman}, Wouter and {Kelly}, Patrick L. and {Mason}, Charlotte A. and {Mercurio}, Amata and {Rosati}, Piero and {Sharon}, Keren and {Trenti}, Michele and {Vulcani}, Benedetta},
        title = "{The Grism Lens-amplified Survey from Space (GLASS). X. Sub-kiloparsec Resolution Gas-phase Metallicity Maps at Cosmic Noon behind the Hubble Frontier Fields Cluster MACS1149.6+2223}",
      journal = {\apj},
         year = 2017,
        month = mar,
       volume = {837},
       number = {1},
          eid = {89},
        pages = {89},
          doi = {10.3847/1538-4357/aa603c},
archivePrefix = {arXiv},
       eprint = {1610.07558},
 primaryClass = {astro-ph.GA},
       adsurl = {https://ui.adsabs.harvard.edu/abs/2017ApJ...837...89W}
}

@ARTICLE{Wang2020,
       author = {{Wang}, Xin and {Jones}, Tucker A. and {Treu}, Tommaso and {Daddi}, Emanuele and {Brammer}, Gabriel B. and {Sharon}, Keren and {Morishita}, Takahiro and {Abramson}, Louis E. and {Colbert}, James W. and {Henry}, Alaina L. and {Hopkins}, Philip F. and {Malkan}, Matthew A. and {Schmidt}, Kasper B. and {Teplitz}, Harry I. and {Vulcani}, Benedetta},
        title = "{A Census of Sub-kiloparsec Resolution Metallicity Gradients in Star-forming Galaxies at Cosmic Noon from HST Slitless Spectroscopy}",
      journal = {\apj},
         year = 2020,
        month = sep,
       volume = {900},
       number = {2},
          eid = {183},
        pages = {183},
          doi = {10.3847/1538-4357/abacce},
archivePrefix = {arXiv},
       eprint = {1911.09841},
 primaryClass = {astro-ph.GA},
       adsurl = {https://ui.adsabs.harvard.edu/abs/2020ApJ...900..183W}
}

@ARTICLE{Wang2022,
       author = {{Wang}, Xin and {Jones}, Tucker and {Vulcani}, Benedetta and {Treu}, Tommaso and {Morishita}, Takahiro and {Roberts-Borsani}, Guido and {Malkan}, Matthew A. and {Henry}, Alaina and {Brammer}, Gabriel and {Strait}, Victoria and {Brada{\v{c}}}, Maru{\v{s}}a and {Boyett}, Kristan and {Calabr{\`o}}, Antonello and {Castellano}, Marco and {Fontana}, Adriano and {Glazebrook}, Karl and {Kelly}, Patrick L. and {Leethochawalit}, Nicha and {Marchesini}, Danilo and {Santini}, P. and {Trenti}, M. and {Yang}, Lilan},
        title = "{Early Results from GLASS-JWST. IV. Spatially Resolved Metallicity in a Low-mass z   3 Galaxy with NIRISS}",
      journal = {\apjl},
         year = 2022,
        month = oct,
       volume = {938},
       number = {2},
          eid = {L16},
        pages = {L16},
          doi = {10.3847/2041-8213/ac959e},
archivePrefix = {arXiv},
       eprint = {2207.13113},
 primaryClass = {astro-ph.GA},
       adsurl = {https://ui.adsabs.harvard.edu/abs/2022ApJ...938L..16W}
}

@ARTICLE{He2024,
       author = {{He}, Xianlong and {Wang}, Xin and {Jones}, Tucker and {Treu}, Tommaso and {Glazebrook}, K. and {Malkan}, Matthew A. and {Vulcani}, Benedetta and {Metha}, Benjamin and {Brada{\v{c}}}, Maru{\v{s}}a and {Brammer}, Gabriel and {Roberts-Borsani}, Guido and {Strait}, Victoria and {Bonchi}, Andrea and {Castellano}, Marco and {Fontana}, Adriano and {Mason}, Charlotte and {Merlin}, Emiliano and {Morishita}, Takahiro and {Paris}, Diego and {Santini}, Paola and {Trenti}, Michele and {Boyett}, Kristan and {Grasha}, K.},
        title = "{Early Results from GLASS-JWST. XXIV. The Mass-Metallicity Relation in Lensed Field Galaxies at Cosmic Noon with NIRISS}",
      journal = {\apjl},
         year = 2024,
        month = jan,
       volume = {960},
       number = {2},
          eid = {L13},
        pages = {L13},
          doi = {10.3847/2041-8213/ad12cd},
archivePrefix = {arXiv},
       eprint = {2312.01932},
 primaryClass = {astro-ph.GA},
       adsurl = {https://ui.adsabs.harvard.edu/abs/2024ApJ...960L..13H}
}

@ARTICLE{LiS2025,
       author = {{Li}, Sijia and {Wang}, Xin and {Chen}, Yuguang and {Jones}, Tucker and {Treu}, Tommaso and {Glazebrook}, Karl and {He}, Xianlong and {Henry}, Alaina and {Meng}, Xiao-Lei and {Morishita}, Takahiro and {Roberts-Borsani}, Guido and {Yang}, Lilan and {Yu}, Hao-Ran and {Calabr{\`o}}, Antonello and {Castellano}, Marco and {Leethochawalit}, Nicha and {Metha}, Benjamin and {Nanayakkara}, Themiya and {Roy}, Namrata and {Vulcani}, Benedetta},
        title = "{Early Results from GLASS-JWST. XXV. Electron Density in the Interstellar Medium at 0.7 {\ensuremath{\lesssim}} z {\ensuremath{\lesssim}} 9.3 with NIRSpec High-resolution Spectroscopy}",
      journal = {\apjl},
         year = 2025,
        month = jan,
       volume = {979},
       number = {1},
          eid = {L13},
        pages = {L13},
          doi = {10.3847/2041-8213/ad9eac},
archivePrefix = {arXiv},
       eprint = {2412.08382},
 primaryClass = {astro-ph.GA},
       adsurl = {https://ui.adsabs.harvard.edu/abs/2025ApJ...979L..13L}
}

@ARTICLE{Marszewski2024,
       author = {{Marszewski}, Andrew and {Sun}, Guochao and {Faucher-Gigu{\`e}re}, Claude-Andr{\'e} and {Hayward}, Christopher C. and {Feldmann}, Robert},
        title = "{The High-Redshift Gas-Phase Mass{\textendash}Metallicity Relation in FIRE-2}",
      journal = {\apjl},
         year = 2024,
        month = jun,
       volume = {967},
       number = {2},
          eid = {L41},
        pages = {L41},
          doi = {10.3847/2041-8213/ad4cee},
archivePrefix = {arXiv},
       eprint = {2403.08853},
 primaryClass = {astro-ph.GA},
       adsurl = {https://ui.adsabs.harvard.edu/abs/2024ApJ...967L..41M}
}

@ARTICLE{Marszewski2025,
       author = {{Marszewski}, Andrew and {Faucher-Gigu{\`e}re}, Claude-Andr{\'e} and {Feldmann}, Robert and {Sun}, Guochao},
        title = "{Explaining the Weak Evolution of the High-Redshift Mass-Metallicity Relation with Galaxy Burst Cycles}",
      journal = {arXiv e-prints},
         year = 2025,
        month = may,
          eid = {arXiv:2505.22712},
        pages = {arXiv:2505.22712},
          doi = {10.48550/arXiv.2505.22712},
archivePrefix = {arXiv},
       eprint = {2505.22712},
 primaryClass = {astro-ph.GA},
       adsurl = {https://ui.adsabs.harvard.edu/abs/2025arXiv250522712M}
}

@ARTICLE{Mannucci2010,
       author = {{Mannucci}, F. and {Cresci}, G. and {Maiolino}, R. and {Marconi}, A. and {Gnerucci}, A.},
        title = "{A fundamental relation between mass, star formation rate and metallicity in local and high-redshift galaxies}",
      journal = {\mnras},
         year = 2010,
        month = nov,
       volume = {408},
       number = {4},
        pages = {2115-2127},
          doi = {10.1111/j.1365-2966.2010.17291.x},
archivePrefix = {arXiv},
       eprint = {1005.0006},
 primaryClass = {astro-ph.CO},
       adsurl = {https://ui.adsabs.harvard.edu/abs/2010MNRAS.408.2115M}
}

@ARTICLE{Bothwell2013,
       author = {{Bothwell}, M.~S. and {Maiolino}, R. and {Kennicutt}, R. and {Cresci}, G. and {Mannucci}, F. and {Marconi}, A. and {Cicone}, C.},
        title = "{A fundamental relation between the metallicity, gas content and stellar mass of local galaxies}",
      journal = {\mnras},
         year = 2013,
        month = aug,
       volume = {433},
       number = {2},
        pages = {1425-1435},
          doi = {10.1093/mnras/stt817},
archivePrefix = {arXiv},
       eprint = {1304.4940},
 primaryClass = {astro-ph.CO},
       adsurl = {https://ui.adsabs.harvard.edu/abs/2013MNRAS.433.1425B}
}

@ARTICLE{Gibson2013,
       author = {{Gibson}, B.~K. and {Pilkington}, K. and {Brook}, C.~B. and {Stinson}, G.~S. and {Bailin}, J.},
        title = "{Constraining sub-grid physics with high-redshift spatially-resolved metallicity distributions}",
      journal = {\aap},
         year = 2013,
        month = jun,
       volume = {554},
          eid = {A47},
        pages = {A47},
          doi = {10.1051/0004-6361/201321239},
archivePrefix = {arXiv},
       eprint = {1304.3020},
 primaryClass = {astro-ph.GA},
       adsurl = {https://ui.adsabs.harvard.edu/abs/2013A&A...554A..47G}
}

@ARTICLE{Hemler2021,
       author = {{Hemler}, Z.~S. and {Torrey}, Paul and {Qi}, Jia and {Hernquist}, Lars and {Vogelsberger}, Mark and {Ma}, Xiangcheng and {Kewley}, Lisa J. and {Nelson}, Dylan and {Pillepich}, Annalisa and {Pakmor}, R{\"u}diger and {Marinacci}, Federico},
        title = "{Gas-phase metallicity gradients of TNG50 star-forming galaxies}",
      journal = {\mnras},
         year = 2021,
        month = sep,
       volume = {506},
       number = {2},
        pages = {3024-3048},
          doi = {10.1093/mnras/stab1803},
archivePrefix = {arXiv},
       eprint = {2007.10993},
 primaryClass = {astro-ph.GA},
       adsurl = {https://ui.adsabs.harvard.edu/abs/2021MNRAS.506.3024H}
}

@ARTICLE{Tissera2022,
       author = {{Tissera}, Patricia B. and {Rosas-Guevara}, Yetli and {Sillero}, Emanuel and {Pedrosa}, Susana E. and {Theuns}, Tom and {Bignone}, Lucas},
        title = "{The evolution of the oxygen abundance gradients in star-forming galaxies in the EAGLE simulations}",
      journal = {\mnras},
         year = 2022,
        month = apr,
       volume = {511},
       number = {2},
        pages = {1667-1684},
          doi = {10.1093/mnras/stab3644},
archivePrefix = {arXiv},
       eprint = {2112.06553},
 primaryClass = {astro-ph.GA},
       adsurl = {https://ui.adsabs.harvard.edu/abs/2022MNRAS.511.1667T}
}

@ARTICLE{Sillero2017,
       author = {{Sillero}, Emanuel and {Tissera}, Patricia B. and {Lambas}, Diego G. and {Michel-Dansac}, Leo},
        title = "{The evolution of the metallicity gradient and the star formation efficiency in disc galaxies}",
      journal = {\mnras},
         year = 2017,
        month = dec,
       volume = {472},
       number = {4},
        pages = {4404-4413},
          doi = {10.1093/mnras/stx2265},
archivePrefix = {arXiv},
       eprint = {1709.00438},
 primaryClass = {astro-ph.GA},
       adsurl = {https://ui.adsabs.harvard.edu/abs/2017MNRAS.472.4404S}
}

@ARTICLE{Aguirre2005,
       author = {{Aguirre}, Anthony and {Schaye}, Joop and {Hernquist}, Lars and {Kay}, Scott and {Springel}, Volker and {Theuns}, Tom},
        title = "{Confronting Cosmological Simulations with Observations of Intergalactic Metals}",
      journal = {\apjl},
         year = 2005,
        month = feb,
       volume = {620},
       number = {1},
        pages = {L13-L17},
          doi = {10.1086/428498},
archivePrefix = {arXiv},
       eprint = {astro-ph/0411076},
 primaryClass = {astro-ph},
       adsurl = {https://ui.adsabs.harvard.edu/abs/2005ApJ...620L..13A}
}

@ARTICLE{Escala2018,
       author = {{Escala}, Ivanna and {Wetzel}, Andrew and {Kirby}, Evan N. and {Hopkins}, Philip F. and {Ma}, Xiangcheng and {Wheeler}, Coral and {Kere{\v{s}}}, Du{\v{s}}an and {Faucher-Gigu{\`e}re}, Claude-Andr{\'e} and {Quataert}, Eliot},
        title = "{Modelling chemical abundance distributions for dwarf galaxies in the Local Group: the impact of turbulent metal diffusion}",
      journal = {\mnras},
         year = 2018,
        month = feb,
       volume = {474},
       number = {2},
        pages = {2194-2211},
          doi = {10.1093/mnras/stx2858},
archivePrefix = {arXiv},
       eprint = {1710.06533},
 primaryClass = {astro-ph.GA},
       adsurl = {https://ui.adsabs.harvard.edu/abs/2018MNRAS.474.2194E}
}

@ARTICLE{Semenov2024,
       author = {{Semenov}, Vadim A.},
        title = "{Capturing Turbulence with Numerical Dissipation: a Simple Dynamical Model for Unresolved Turbulence in Hydrodynamic Simulations}",
      journal = {arXiv e-prints},
         year = 2024,
        month = oct,
          eid = {arXiv:2410.23339},
        pages = {arXiv:2410.23339},
          doi = {10.48550/arXiv.2410.23339},
archivePrefix = {arXiv},
       eprint = {2410.23339},
 primaryClass = {astro-ph.GA},
       adsurl = {https://ui.adsabs.harvard.edu/abs/2024arXiv241023339S}
}

@ARTICLE{Baker2025,
       author = {{Baker}, William M. and {Tacchella}, Sandro and {Johnson}, Benjamin D. and {Nelson}, Erica and {Suess}, Katherine A. and {D'Eugenio}, Francesco and {Curti}, Mirko and {de Graaff}, Anna and {Ji}, Zhiyuan and {Maiolino}, Roberto and {Robertson}, Brant and {Scholtz}, Jan and {Alberts}, Stacey and {Arribas}, Santiago and {Boyett}, Kristan and {Bunker}, Andrew J. and {Carniani}, Stefano and {Charlot}, Stephane and {Chen}, Zuyi and {Chevallard}, Jacopo and {Curtis-Lake}, Emma and {Danhaive}, A. Lola and {DeCoursey}, Christa and {Egami}, Eiichi and {Eisenstein}, Daniel J. and {Endsley}, Ryan and {Hausen}, Ryan and {Helton}, Jakob M. and {Kumari}, Nimisha and {Looser}, Tobias J. and {Maseda}, Michael V. and {Pusk{\'a}s}, D{\'a}vid and {Rieke}, Marcia and {Sandles}, Lester and {Sun}, Fengwu and {{\"U}bler}, Hannah and {Williams}, Christina C. and {Willmer}, Christopher N.~A. and {Witstok}, Joris},
        title = "{A core in a star-forming disc as evidence of inside-out growth in the early Universe}",
      journal = {Nature Astronomy},
         year = 2025,
        month = jan,
       volume = {9},
        pages = {141-154},
          doi = {10.1038/s41550-024-02384-8},
archivePrefix = {arXiv},
       eprint = {2306.02472},
 primaryClass = {astro-ph.GA},
       adsurl = {https://ui.adsabs.harvard.edu/abs/2025NatAs...9..141B}
}

@ARTICLE{Luo2024,
       author = {{Luo}, Xiong and {Wang}, Huiyuan and {Cui}, Weiguang and {Mo}, Houjun and {Li}, RenJie and {Jing}, Yipeng and {Katz}, Neal and {Dav{\'e}}, Romeel and {Yang}, Xiaohu and {Chen}, Yangyao and {Li}, Hao and {Huang}, Shuiyao},
        title = "{ELUCID. VIII. Simulating the Coma Galaxy Cluster to Calibrate Model and Understand Feedback}",
      journal = {\apj},
         year = 2024,
        month = may,
       volume = {966},
       number = {2},
          eid = {236},
        pages = {236},
          doi = {10.3847/1538-4357/ad392e},
archivePrefix = {arXiv},
       eprint = {2401.14730},
 primaryClass = {astro-ph.GA},
       adsurl = {https://ui.adsabs.harvard.edu/abs/2024ApJ...966..236L}
}

@ARTICLE{Mo20241,
       author = {{Mo}, Houjun and {Chen}, Yangyao and {Wang}, Huiyuan},
        title = "{A two-phase model of galaxy formation: I. The growth of galaxies and supermassive black holes}",
      journal = {\mnras},
         year = 2024,
        month = aug,
       volume = {532},
       number = {4},
        pages = {3808-3838},
          doi = {10.1093/mnras/stae1727},
archivePrefix = {arXiv},
       eprint = {2311.05030},
 primaryClass = {astro-ph.GA},
       adsurl = {https://ui.adsabs.harvard.edu/abs/2024MNRAS.532.3808M}
}

@ARTICLE{Chen20253,
       author = {{Chen}, Yangyao and {Mo}, Houjun and {Wang}, Huiyuan},
        title = "{A two-phase model of galaxy formation: III. The formation of globular clusters}",
      journal = {\mnras},
         year = 2025,
        month = jun,
       volume = {540},
       number = {1},
        pages = {1235-1271},
          doi = {10.1093/mnras/staf791},
archivePrefix = {arXiv},
       eprint = {2405.18735},
 primaryClass = {astro-ph.GA},
       adsurl = {https://ui.adsabs.harvard.edu/abs/2025MNRAS.540.1235C}
}

@ARTICLE{Chen20242,
       author = {{Chen}, Yangyao and {Mo}, Houjun and {Wang}, Huiyuan},
        title = "{A two-phase model of galaxy formation - II. The size-mass relation of dynamically hot galaxies}",
      journal = {\mnras},
         year = 2024,
        month = aug,
       volume = {532},
       number = {4},
        pages = {4340-4349},
          doi = {10.1093/mnras/stae1757},
archivePrefix = {arXiv},
       eprint = {2311.11713},
 primaryClass = {astro-ph.GA},
       adsurl = {https://ui.adsabs.harvard.edu/abs/2024MNRAS.532.4340C}
}

@ARTICLE{Liang2025,
       author = {{Liang}, Jinning and {Jiang}, Fangzhou and {Mo}, Houjun and {Benson}, Andrew and {Dekel}, Avishai and {Tavron}, Noa and {Hopkins}, Philip F. and {Ho}, Luis C.},
        title = "{Connection between galaxy morphology and dark-matter halo structure I: a running threshold for thin discs and size predictors from the dark sector}",
      journal = {\mnras},
         year = 2025,
        month = aug,
       volume = {541},
       number = {3},
        pages = {2304-2323},
          doi = {10.1093/mnras/staf947},
archivePrefix = {arXiv},
       eprint = {2403.14749},
 primaryClass = {astro-ph.GA},
       adsurl = {https://ui.adsabs.harvard.edu/abs/2025MNRAS.541.2304L}
}

@ARTICLE{He2024S,
       author = {{He}, Zhicheng and {Chen}, Zhifu and {Liu}, Guilin and {Wang}, Tinggui and {Ho}, Luis C. and {Wang}, Junxian and {Bian}, Weihao and {Cai}, Zheng and {Mou}, Guobin and {Gu}, Qiusheng and {Wang}, Zhiwen},
        title = "{The transition from galaxy-wide gas inflow to outflow in quasar host galaxies}",
      journal = {Science China Physics, Mechanics, and Astronomy},
         year = 2024,
        month = dec,
       volume = {67},
       number = {12},
          eid = {129512},
        pages = {129512},
          doi = {10.1007/s11433-024-2475-7},
archivePrefix = {arXiv},
       eprint = {2408.04458},
 primaryClass = {astro-ph.GA},
       adsurl = {https://ui.adsabs.harvard.edu/abs/2024SCPMA..6729512H}
}

@ARTICLE{Jin2024a,
       author = {{Jin}, Bingcheng and {Ho}, Luis C. and {Sun}, Wen},
        title = "{A High Incidence of Central Star Formation Inferred from the Color Gradients of Galaxies at $z>4$}",
      journal = {arXiv e-prints},
         year = 2024,
        month = dec,
          eid = {arXiv:2412.03455},
        pages = {arXiv:2412.03455},
          doi = {10.48550/arXiv.2412.03455},
archivePrefix = {arXiv},
       eprint = {2412.03455},
 primaryClass = {astro-ph.GA},
       adsurl = {https://ui.adsabs.harvard.edu/abs/2024arXiv241203455J}
}

@ARTICLE{Peng2025P,
       author = {{Peng}, Yue-Chang and {Wang}, Jian-Min and {Zhao}, Yu and {Ho}, Luis C.},
        title = "{Bulge Oscillation Driven by Outflows of Active Galactic Nuclei. I. Fast Outflow Case}",
      journal = {\apj},
         year = 2025,
        month = feb,
       volume = {980},
       number = {1},
          eid = {22},
        pages = {22},
          doi = {10.3847/1538-4357/ada275},
archivePrefix = {arXiv},
       eprint = {2412.17725},
 primaryClass = {astro-ph.GA},
       adsurl = {https://ui.adsabs.harvard.edu/abs/2025ApJ...980...22P}
}

@ARTICLE{Bassini20231,
       author = {{Bassini}, Luigi and {Feldmann}, Robert and {Gensior}, Jindra and {Hayward}, Christopher C. and {Faucher-Gigu{\`e}re}, Claude-Andr{\'e} and {Cenci}, Elia and {Liang}, Lichen and {Bernardini}, Mauro},
        title = "{The inefficiency of stellar feedback in driving galactic outflows in massive galaxies at high redshift}",
      journal = {\mnras},
         year = 2023,
        month = nov,
       volume = {525},
       number = {4},
        pages = {5388-5405},
          doi = {10.1093/mnras/stad2617},
archivePrefix = {arXiv},
       eprint = {2211.08423},
 primaryClass = {astro-ph.GA},
       adsurl = {https://ui.adsabs.harvard.edu/abs/2023MNRAS.525.5388B}
}

@BOOK{Mo2010,
       author = {{Mo}, Houjun and {van den Bosch}, Frank C. and {White}, Simon},
        title = "{Galaxy Formation and Evolution}",
         year = 2010,
          doi = {10.1017/CBO9780511807244},
       adsurl = {https://ui.adsabs.harvard.edu/abs/2010gfe..book.....M}
}

@ARTICLE{LiTie2025,
       author = {{Li}, Tie and {Zhang}, Hong-Xin and {Lyu}, Wenhe and {Tang}, Yimeng and {Yao}, Yao and {Wang}, Enci and {Rong}, Yu and {Chen}, Guangwen and {Kong}, Xu and {Bian}, Fuyan and {Gu}, Qiusheng and {Johnston}, Evelyn J. and {Li}, Xin and {Mao}, Shude and {Shi}, Yong and {Wang}, Junfeng and {Wang}, Xin and {Yu}, Xiaoling and {Zheng}, Zhiyuan},
        title = "{A negative stellar mass‑gaseous metallicity gradient relation of dwarf galaxies modulated by stellar feedback}",
      journal = {\aap},
         year = 2025,
        month = jun,
       volume = {698},
          eid = {A208},
        pages = {A208},
          doi = {10.1051/0004-6361/202452978},
archivePrefix = {arXiv},
       eprint = {2504.17541},
 primaryClass = {astro-ph.GA},
       adsurl = {https://ui.adsabs.harvard.edu/abs/2025A&A...698A.208L}
}

@ARTICLE{Lyu2025,
       author = {{Lyu}, Cheqiu and {Wang}, Enci and {Zhang}, Hongxin and {Peng}, Yingjie and {Wang}, Xin and {Li}, Haixin and {Ma}, Chengyu and {Yu}, Haoran and {Chen}, Zeyu and {Jia}, Cheng and {Kong}, Xu},
        title = "{Dominant Role of Coplanar Inflows in Driving Disk Evolution Revealed by Gas-phase Metallicity Gradients}",
      journal = {\apjl},
         year = 2025,
        month = mar,
       volume = {981},
       number = {1},
          eid = {L6},
        pages = {L6},
          doi = {10.3847/2041-8213/adb4ed},
archivePrefix = {arXiv},
       eprint = {2502.12409},
 primaryClass = {astro-ph.GA},
       adsurl = {https://ui.adsabs.harvard.edu/abs/2025ApJ...981L...6L}
}

@ARTICLE{Andrews2013,
       author = {{Andrews}, Brett H. and {Martini}, Paul},
        title = "{The Mass-Metallicity Relation with the Direct Method on Stacked Spectra of SDSS Galaxies}",
      journal = {\apj},
         year = 2013,
        month = mar,
       volume = {765},
       number = {2},
          eid = {140},
        pages = {140},
          doi = {10.1088/0004-637X/765/2/140},
archivePrefix = {arXiv},
       eprint = {1211.3418},
 primaryClass = {astro-ph.CO},
       adsurl = {https://ui.adsabs.harvard.edu/abs/2013ApJ...765..140A}
}
\bibliographystyle{aasjournal}

\end{CJK*}
\end{document}